\documentclass[letter,10pt]{article}

\usepackage[margin=1in]{geometry}
\usepackage{amsmath,amssymb}
\usepackage{changepage}
\usepackage[utf8x]{inputenc}
\usepackage{textcomp,marvosym}
\usepackage{cite}
\usepackage{nameref}
\usepackage[hidelinks]{hyperref}
\usepackage[right]{lineno}
\usepackage{microtype}
\usepackage[table]{xcolor}
\usepackage{array}
\usepackage{booktabs}
\usepackage{tabularx}
\usepackage{multirow}
\usepackage{bbm}
\usepackage{pgf}
\usepackage{tikz}
\usepackage{subcaption}
\usepackage{enumitem}
\usepackage{algpseudocode}
\usepackage{algorithm}
\usepackage{float}

\newcolumntype{L}[1]{>{\raggedright\arraybackslash}p{#1}}

\let\pgfimageWithoutPath\pgfimage
\renewcommand{\pgfimage}[2][]{\pgfimageWithoutPath[#1]{figures/#2}}

\newenvironment{affiliations}{%
    \setcounter{enumi}{1}%
    \setlength{\parindent}{0in}%
    \slshape\sloppy%
    \begin{list}{\upshape$^{\arabic{enumi}}$}{%
        \usecounter{enumi}%
        \setlength{\leftmargin}{0in}%
        \setlength{\topsep}{0in}%
        \setlength{\labelsep}{0in}%
        \setlength{\labelwidth}{0in}%
        \setlength{\listparindent}{0in}%
        \setlength{\itemsep}{0ex}%
        \setlength{\parsep}{0in}%
        }
    }{\end{list}\par\vspace{12pt}}

\title{HNCcorr: A Novel Combinatorial Approach for Cell Identification in
	Calcium-Imaging Movies}
\author{Quico Spaen$^{1}$, Dorit S. Hochbaum$^1$, Roberto As\'in-Ach\'a$^2$}
\date{}

\begin{document}

\maketitle

\begin{affiliations}
 \item Department of Industrial Engineering \& Operations Research, University of California, Berkeley, Etcheverry Hall, CA 94720, USA
 \item Department of Computer Science, Universidad de Concepci\'on, Concepci\'on, Chile
\end{affiliations}

	\section*{Abstract}
Calcium imaging has emerged as a workhorse method in neuroscience to investigate patterns of neuronal activity. Instrumentation to acquire calcium imaging movies has rapidly progressed and has become standard across labs. Still, algorithms to automatically detect and extract activity signals from calcium imaging movies are highly variable from~lab~to~lab and more advanced algorithms are continuously being developed. Here we present HNCcorr, a novel algorithm for cell identification in calcium imaging movies based on combinatorial optimization. The algorithm identifies cells by finding distinct groups of highly similar pixels in correlation space, where a pixel is represented by the vector of correlations to a set of other pixels. The HNCcorr algorithm achieves the best known results for the cell identification benchmark of Neurofinder, and guarantees an optimal solution to the underlying deterministic optimization model resulting in a transparent mapping from input data to outcome.

	\section*{Introduction}

In the past decade new methods to measure neuronal activity have emerged based on optical imaging of activity-dependent sensors \cite{HamGrePar15,ScaHau09}.
In particular, optical imaging using genetically-encoded calcium indicators, known as calcium imaging, has become a key tool for circuits and systems neuroscience and is currently used by hundreds of labs across the world \cite{HamGrePar15}. While instrumentation and the calcium indicators have matured rapidly, algorithms to automatically detect and extract activity signals from calcium imaging movies are highly variable from~lab~to~lab. More advanced algorithms are continuously being developed.  As a result, the analysis of calcium imaging movies is non-standardized and requires extensive human supervision \cite{HamGrePar15,PerCheSvo15}. 

Calcium imaging movies collected during an experiment require post-processing to extract the fluorescence signals of the individual cells in the movie. This process consists of two steps. First, in the {\em cell identification problem}, the goal is to identify the {\em spatial footprint} of the cells in the movie. The spatial footprint of a cell is the set of pixels that contain this cell.
Then, in the {\em signal extraction} step, the fluorescence signal of each cell is extracted based on the intensities of the pixels in the spatial footprint of the cell.
For the signal extraction step a number of effective algorithms have been proposed~\cite{VogPacMac10,GreLanKas10,TheBerFro15,PneMerPak13}.
For the cell identification problem, in addition to manual identification of the cells, two types of automated procedures have been proposed:
Shape-based segmentation algorithms~\cite{PacPacPet13,neuNet16}, and signal-based algorithms. The signal-based algorithm use one of two techniques:
Matrix factorization ~\cite{MukNimSch09,MarMaeMor14,PnePan13,PneMacGro13,DieHam14,PneSouGao16,PacStrSch16} or graph partitioning~\cite{KaiZarDan14}. 
See the literature review for a detailed discussion of these methods.

In this paper we present HNCcorr, a new signal-based method for cell identification in calcium imaging movies.
Our method is novel in that it identifies the spatial footprint of a cell by mapping the pixels to {\emph correlation space}, where a pixel is represented by the vector of correlations to a set of other pixels, and the use of a graph-partitioning model that can be solved optimally.
In contrast, all other techniques rely on heuristics that do not guarantee optimal solutions for the underlying models.
By mapping the pixels to correlation space, the pixels of each cell group together and also the background pixels congregate.
HNCcorr then uses the {\emph HNC} model~\cite{Hoc10,Hoc13} to identify cells one at a time by finding a set of pixels that are close in correlation space, but whose correlation patterns are distinct from the rest of the pixels.
This type of model originates in image segmentation and is also effective in general data mining~\cite{Hoc10,Hoc13,HocHsuYan12,HocLyuBer13,HocBau16,BauHocSpa16}. 
This model is solved optimally with a fast combinatorial optimization algorithm.
A Python implementation of HNCcorr is available on GitHub: \url{https://github.com/quic0/HNCcorr}.

Experimentally, the HNCcorr outperforms other state-of-the-art approaches in the Neurofinder challenge \cite{NeuFinder16}. 
It is able to accurately detect more cells than competing algorithms. 
When combined with the shape-based algorithm Donuts~\cite{PacPacPet13} for detecting cells without a signal, HNCcorr achieves the best known average score for the Neurofinder benchmark as of January 2017.




	\section*{Literature Review}
\label{sec:litReview}

The known approaches for the cell identification problem can be classified into two different categories: Shape-based algorithms and signal-based algorithms.
The algorithms in the first category identify the cell bodies by looking for the shape of a cell in an input image, which is typically the image containing the average intensity of each pixel over time.
The second category of algorithms identifies cells by identifying clusters of pixels with a similar fluorescence signal over time.
The signal-based algorithms can be split into two types of techniques: Matrix factorization and graph partitioning.
Next, each of these techniques is discussed in detail.

\begin{description}
  \item[Shape-based identification:] Pachitariu et al.~\cite{PacPacPet13} propose a generative model that aims to capture regularities in an image composed of many elements of few different types.
  The model assumes that a shape (i.e. a cell body) is built from a set of learned features (e.g. edges, curved edges, corners, etc).
  Based on the learned features, the model can predict the probability that a group of pixels represents a cell body.
  The algorithm Donuts uses these learned features to identify a set of shapes that should correspond with cell bodies.

  More recently, an algorithm named Conv2d that applies convolutional neural networks in a two-dimensional image has been proposed \cite{neuNet16}.
  The network is trained on data where the cell bodies have been identified. Based on this trained network, it predicts the probability that each pixel belongs to a cell and constructs the spatial footprints of the cells.

  Both of these approaches tend to work well whenever the cell body is easily distinguishable in the input image.
  However, cells that are not clearly visible in this image cannot be found.
  In practice, due to the acquisition technique, many cells are only visible in the small number of frames where they are activated.
  These cells do not show in the image containing the average intensity of each pixel and can thus not be identified.
  Furthermore, these techniques are also not well-suited for detecting overlapping cells, since these cells can only be separated based on their signals.

  \item[Matrix factorization:] Matrix factorization is a set of techniques that factorize a matrix as a product of multiple low-dimensional matrices. These techniques provide the spatial footprint and the signals of the cells. Mukamel et al.~\cite{MukNimSch09} was the first paper that introduced the idea that the intensity of a pixel over time is a noisy measurement of a mixture of signals.
  The movie can be split into a set of signals, each having a spatial and a temporal component.
  The spatial component measures the extent to which a signal contributes to the intensity of each of the pixels and the temporal component represents the value of the signal over time.
  To extract the cells' signals from the movie, the method first uses Principal Component Analysis (PCA) to reduce the dimensions of the data and noise.
  Then, the signals are identified using spatio-temporal independent component analysis (ICA).

  In recent work~\cite{DieHam14,MarMaeMor14,PnePan13}, it was recognized that this idea can be formalized as matrix factorization.
  They consider the following generative model for the signal of the intensities of the pixels:
  $$
  	F = A C + B + \text{i.i.d. noise}
  $$
  where $F$ is a non-negative matrix containing the fluorescence of all pixels over time, $A$ is a non-negative matrix that measures the contribution of each cell's signal to the intensity of each pixel, $C$ is a matrix that captures the signal of the cells at each time step, and $B$ is a background signal.
  The matrix $F$ is known and we wish to infer the values of the matrices $A,C,B$, where $A$ and $C$ respectively capture the spatial and temporal components of each of the cells.
  The matrices are inferred by minimizing the objective function $\| F - AC - B \|^2 +\Omega(A,B,C)$, where $\| \cdot \|$ is a matrix norm and $\Omega(A,B,C)$ is a regularization function on the matrices $A$, $B$, and/or $C$.
  An iterative procedure known as coordinate descent (a special case is Alternating Least Squares) is applied to this problem.
  Starting from an initial solution, the iterative algorithm alternates between fixing $A$ and optimizing over $C$ and fixing $C$ and optimizing over $A$ until the solution converges.
  All of the papers listed above present a variant of this approach.
  Table~$\ref{tab:overviewMFalgo}$ summarizes their differences.

  \begin{table}
    \footnotesize
  	\caption{Overview of the specific properties used in each matrix factorization algorithm. See corresponding papers for more detail.}\label{tab:overviewMFalgo}
  	\begin{tabularx}{\textwidth}{L{2.5cm}L{2.8cm}L{2.6cm}L{1.3cm}XL{1.5cm}}
  		\toprule
  		Paper & Assumptions on temporal component matrix $C$ & Assumptions on background matrix $B$ & Objective norm & Regularization &  Initialization \\
  		\cmidrule(r){1-1} \cmidrule(lr){2-2} \cmidrule(lr){3-3} \cmidrule(lr){4-4} \cmidrule(lr){5-5} \cmidrule(l){6-6}
  		Maruyama et al~\cite{MarMaeMor14} & None & $ b\times f $ - Rank 1 matrix & $\ell_2$ norm & None & Random \\
  		Pnevmatikakis et al.~\cite{PnePan13,PneMacGro13} & AR($p$) process with spikes &  $b\, \mathbbm{1}^T$  - Constant for each pixel & $\ell_2$ norm & Bayesian prior on sparsity of spikes and nuclear norm penalty on sparsity of $A$ & Clustering method \\
  		Diego-Andilla \& Hamprecht~\cite{DieHam14} & Sparse low-rank matrix & $ b\times f $ -  Rank 1 matrix & $ \ell_2$ norm & Total variation norm on the spatial component of the background $b$ & Random \\
  		Pnevmatikakis et al.~\cite{PneGaoSou14,PneSouGao16} & AR($p$) process with spikes & $b\times f $ -  Rank 1 matrix & $\ell_1$ norm & Constraints on variance of noise vector & Greedy algorithm\\
  		Pachitariu et al.~\cite{PacStrSch16} & None & Set of isotropic 2d raised cosine functions & $\ell_2$ norm & None & Not specified\\
  		\bottomrule
  	\end{tabularx}
  \end{table}

  Although these techniques claim adequate results, the models and their solution techniques suffer from major shortcomings. The optimization problem of matrix factorization is inherently non-convex and computationally intractable, which means it cannot be solved optimally.
  Instead, the associated algorithms are heuristics that only find a local minimum close to the initial solution.
  As a result, the quality of the solution is highly dependent on the initial solution and may not generalize across datasets.

  Furthermore, these algorithms are based on a set of assumptions on the dynamics of the  fluorescent calcium indicator.
  These assumptions do not necessarily capture the actual dynamics.
  In contrast, the HNCcorr algorithm makes no specific assumptions about the underlying dynamics of the fluorescent indicator.

  \item[Graph partitioning:] This group of techniques identifies cells by partitioning a graph in which the nodes correspond to the pixels and the edges measures the similarity between pairs of pixels.
  In \cite{KaiZarDan14}, Kaifosh et al. propose a model in which they find the ROIs, consisting of one or more cells, by repeatedly subdividing the set of pixels into coherent groups.
  To do this, they use an eigenvector-based heuristic for an optimization problem known as Normalized Cut (NC)~\cite{ShiMal00}.
  This approach suffers from two major drawbacks.
  First, the Normalized Cut model is not appropriate for this problem (see the section on methodology for details).
  Second, they rely on a heuristic for solving the optimization problem, which may lead to suboptimal results.

  The HNCcorr algorithm is similar to the algorithm of~\cite{KaiZarDan14} in that it partitions the pixels based on a similarity graph.
  However, the algorithm uses both a different graph and a different optimization model.
  As a result, it is not affected by the shortcomings listed above.
\end{description}

In contrast to the existing approaches, the method proposed here solves a discrete optimization problem and guarantees an optimal solution.
The algorithm is robust in that it does not rely on initialization and requires minimal parameter tuning.
Furthermore, our algorithm uses the novel idea to identify cell bodies based on the similarity between pixels in correlation space.

	\section*{Methodology}
HNCcorr aggregates a set of pixels in a cluster so they are highly similar to each other, and highly non-similar to the pixels not in the cluster.  These clusters form the spatial footprints of cells.  HNCcorr identifies these clusters one at a time.
It is noted that the intensity of a pixel in the spatial footprint of a cell is a noisy observation of the signal of the cell over time.
This implies that pixels in the spatial footprint of a cell should have the same intensity pattern over time, and that the intensities of these pixels should be highly correlated.  To that end, we associate with each pixel the vector of signal correlations with other pixels in the window, referred to as the {\em correlation space}. The distances between those vectors are used in HNCcorr as the similarity measure for the clustering problem.
Next, we first describe the core components of the HNCcorr algorithm in detail.

\subsection*{Clustering pixels with HNC}
The pixels in the spatial footprint of a cell have highly similar signals.
We use the HNC model~\cite{Hoc10,Hoc13} to identify the clusters of pixels in a graph.
This graph is an undirected graph $G= (V,E)$ where the node set $V$ represents the pixels and the edges in set $E$ represent similarity relations between pixels.
With each edge $[i,j] \in E$, we associate a similarity weight $w_{ij} \in [0,1]$ where $w_{ij}$ is close to $1$ if pixels $i,j$ are highly similar and close to $0$ if they are not.

The goal in the HNC model is to find a partition of the nodes (pixels) into the set $S$, which represents the spatial footprint of a single cell, and the set $\bar{S} = V \setminus S$, which are the remaining pixels, so that the similarity between pixels in $S$ and the pixels in $\bar{S}$ is as low as possible, low {\em inter-similarity}, and the pixels in the spatial footprint $S$ are highly similar to each other, high {\em intra-similarity}.  This is expresses as the following optimization problem:
\begin{equation}
\min_{ \emptyset \subset S \subset V } \overbrace{\sum_{\substack{[i,j]\in E,\\ i \in S,\,j \in \bar{S}}}w_{ij}}^{\text{Inter-similarity}}
- \lambda \overbrace{\sum_{\substack{[i,j]\in E,\\ i \in S,\, j \in S}}w_{ij}}^{\text{Intra-similarity}}. \tag{HNC} \label{eq:HNC}
\end{equation}
The parameter $\lambda \geq 0$ determines the tradeoff between the inter-similarity and the intra-similarity.
Note that minimizing the negative of the intra-similarity term is equivalent to maximizing it.

HNC is closely related to a well-known optimization problem named Normalized Cut (NC) from the field of image segmentation~\cite{ShiMal00,ShaGalSha06}. NC has been previously used for calcium imaging movies to find regions that contain one or more cells by Kaifosh et al.~\cite{KaiZarDan14}.
NC is not an appropriate model for the cell identification problem because it would try to assure that the remaining pixels (i.e. those in $\bar{S}$) are also highly similar.
Furthermore, NC is known to be NP-hard which means that it is computationally intractable.
In practice, it is solved with a heuristic based on eigenvectors~\cite{ShiMal00}.

In contrast, very efficient algorithms exist to optimally solve the HNC problem.
These algorithms~\cite{Hoc10,Hoc13} arise from reducing the HNC problem to the well-known minimum cut/maximum flow problem.
These algorithms are especially powerful in that they guarantee to find an optimal set $S$.
This is a major advantage since it creates a unique and transparent mapping from input to output.
As a result, it is straightforward to understand how changing the input affects the output of the algorithm.
Furthermore, we can use parametric cut/flow algorithms for the minimum cut problem~\cite{GolTar88,Hoc08} that find the optimal solutions for all values of $\lambda$ at once, removing the need for tuning the $\lambda$ parameter.

\subsection*{Seed and window selection}
At each iteration the goal is to segment a single cell not previously identified.
To ensure that the model identifies the correct cell, HNC needs as input a small set of representative pixels, seeds, for $S$ and $\bar{S}$.  For $S$ we select as seed a superpixel, a small square of $k\times k$ pixels where $k\leq 5$.
Note that when $k=1$ a superpixel is a singleton pixel.
We refer to the selected superpixel for $S$ as \emph{the seed}.

For $\bar{S}$ we select a small set of pixels, referred to as the \emph{negative seeds}.
These pixels are picked uniformly from a circle of a sufficiently large radius centered at the seed, such that the cell to be identified, if exists, can be assumed to be contained in the circle.

For each cell to be segmented, only the pixels within a certain vicinity of the seed need to be considered.
For that purpose a sufficiently large {\em window} is constructed around the seed.
The window is an $n_1\times n_2$ rectangular subregion of the movie that contains the seed at its center.
The HNC clustering problem is restricted to the graph defined on the pixels in this window.

To find all cells, one can simply try all possible (super)pixels of a given size as seeds.
Since the procedure for identifying a single cell is computationally very efficient, this exhaustive search is possible.
Nevertheless, this procedure performs a significant amount of unnecessary computation.
For example, selecting two seeds close to each other is likely to give the same outcome.
Hence, it is sufficient to consider at most a few pixels per small subregion of the movie.
Furthermore, a good indication of whether the seed is part of a cell is to compute the average correlation between the seed and the surrounding neighborhood of e.g. $5 \times 5$ pixels. In most cases, it would be sufficient to try only the $30-40$ percent of pixels with the highest average correlation as seeds to detect nearly all of the active cells. Finally, pixels that are part of a previously found cell should be excluded to prevent the same cell from being segmented multiple times.


\subsection*{Measuring similarity: Correlation space}
A standard technique for measuring similarity between signals is to measure the correlation between them.
However, this is not an appropriate measure for the similarity measure $w_{ij}$.
The problem with correlation is that pixels that are not part of any cell are not sufficiently correlated with each other.
As a result, these pixels are as likely to be clustered in $S$ (the spatial footprint) as in $\bar{S}$.
To overcome this, we introduce correlation space.

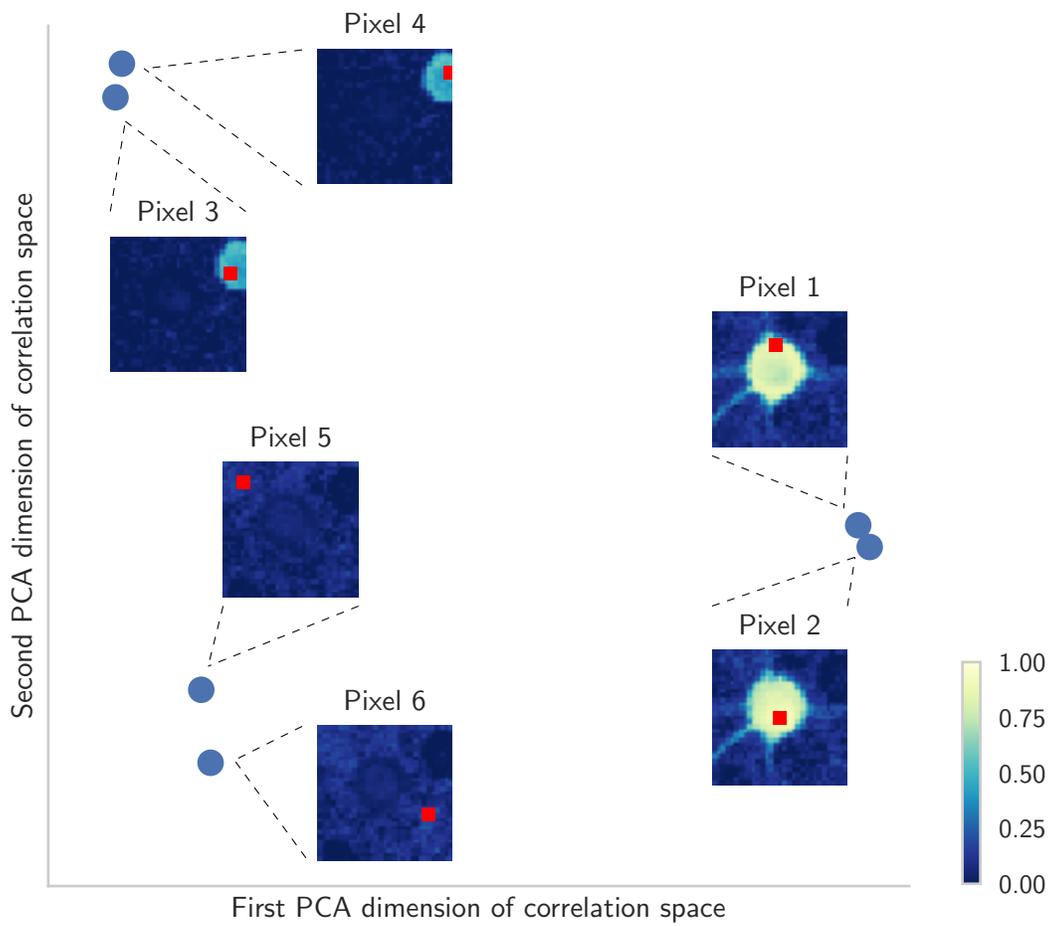
\begin{figure}
	\centering
	\begin{minipage}{\textwidth}
		\centering
		\begin{tikzpicture}[ x = 0.5cm, y=0.5cm]
		\node (corrspace) (0,0) {
\begingroup%
\makeatletter%
\begin{pgfpicture}%
\pgfpathrectangle{\pgfpointorigin}{\pgfqpoint{5.000000in}{5.000000in}}%
\pgfusepath{use as bounding box, clip}%
\begin{pgfscope}%
\pgfsetbuttcap%
\pgfsetmiterjoin%
\definecolor{currentfill}{rgb}{1.000000,1.000000,1.000000}%
\pgfsetfillcolor{currentfill}%
\pgfsetlinewidth{0.000000pt}%
\definecolor{currentstroke}{rgb}{1.000000,1.000000,1.000000}%
\pgfsetstrokecolor{currentstroke}%
\pgfsetdash{}{0pt}%
\pgfpathmoveto{\pgfqpoint{0.000000in}{0.000000in}}%
\pgfpathlineto{\pgfqpoint{5.000000in}{0.000000in}}%
\pgfpathlineto{\pgfqpoint{5.000000in}{5.000000in}}%
\pgfpathlineto{\pgfqpoint{0.000000in}{5.000000in}}%
\pgfpathclose%
\pgfusepath{fill}%
\end{pgfscope}%
\begin{pgfscope}%
\pgfsetbuttcap%
\pgfsetmiterjoin%
\definecolor{currentfill}{rgb}{1.000000,1.000000,1.000000}%
\pgfsetfillcolor{currentfill}%
\pgfsetlinewidth{0.000000pt}%
\definecolor{currentstroke}{rgb}{0.000000,0.000000,0.000000}%
\pgfsetstrokecolor{currentstroke}%
\pgfsetstrokeopacity{0.000000}%
\pgfsetdash{}{0pt}%
\pgfpathmoveto{\pgfqpoint{0.344444in}{0.344444in}}%
\pgfpathlineto{\pgfqpoint{4.850000in}{0.344444in}}%
\pgfpathlineto{\pgfqpoint{4.850000in}{4.850000in}}%
\pgfpathlineto{\pgfqpoint{0.344444in}{4.850000in}}%
\pgfpathclose%
\pgfusepath{fill}%
\end{pgfscope}%
\begin{pgfscope}%
\definecolor{textcolor}{rgb}{0.150000,0.150000,0.150000}%
\pgfsetstrokecolor{textcolor}%
\pgfsetfillcolor{textcolor}%
\pgftext[x=2.597222in,y=0.288889in,,top]{\color{textcolor}\sffamily\fontsize{10.500000}{12.600000}\selectfont First PCA dimension of correlation space}%
\end{pgfscope}%
\begin{pgfscope}%
\definecolor{textcolor}{rgb}{0.150000,0.150000,0.150000}%
\pgfsetstrokecolor{textcolor}%
\pgfsetfillcolor{textcolor}%
\pgftext[x=0.288889in,y=2.597222in,,bottom,rotate=90.000000]{\color{textcolor}\sffamily\fontsize{10.500000}{12.600000}\selectfont Second PCA dimension of correlation space}%
\end{pgfscope}%
\begin{pgfscope}%
\pgfpathrectangle{\pgfqpoint{0.344444in}{0.344444in}}{\pgfqpoint{4.505556in}{4.505556in}} %
\pgfusepath{clip}%
\pgfsetroundcap%
\pgfsetroundjoin%
\pgfsetlinewidth{0.000000pt}%
\definecolor{currentstroke}{rgb}{0.298039,0.447059,0.690196}%
\pgfsetstrokecolor{currentstroke}%
\pgfsetdash{}{0pt}%
\pgfpathmoveto{\pgfqpoint{4.643870in}{2.120442in}}%
\pgfpathlineto{\pgfqpoint{4.583697in}{2.234294in}}%
\pgfpathlineto{\pgfqpoint{0.696275in}{4.475852in}}%
\pgfpathlineto{\pgfqpoint{0.730036in}{4.652099in}}%
\pgfpathlineto{\pgfqpoint{1.145169in}{1.372517in}}%
\pgfpathlineto{\pgfqpoint{1.194110in}{0.990233in}}%
\pgfusepath{}%
\end{pgfscope}%
\begin{pgfscope}%
\pgfpathrectangle{\pgfqpoint{0.344444in}{0.344444in}}{\pgfqpoint{4.505556in}{4.505556in}} %
\pgfusepath{clip}%
\pgfsetbuttcap%
\pgfsetroundjoin%
\definecolor{currentfill}{rgb}{0.298039,0.447059,0.690196}%
\pgfsetfillcolor{currentfill}%
\pgfsetlinewidth{0.000000pt}%
\definecolor{currentstroke}{rgb}{0.298039,0.447059,0.690196}%
\pgfsetstrokecolor{currentstroke}%
\pgfsetdash{}{0pt}%
\pgfsys@defobject{currentmarker}{\pgfqpoint{-0.069444in}{-0.069444in}}{\pgfqpoint{0.069444in}{0.069444in}}{%
\pgfpathmoveto{\pgfqpoint{0.000000in}{-0.069444in}}%
\pgfpathcurveto{\pgfqpoint{0.018417in}{-0.069444in}}{\pgfqpoint{0.036082in}{-0.062127in}}{\pgfqpoint{0.049105in}{-0.049105in}}%
\pgfpathcurveto{\pgfqpoint{0.062127in}{-0.036082in}}{\pgfqpoint{0.069444in}{-0.018417in}}{\pgfqpoint{0.069444in}{0.000000in}}%
\pgfpathcurveto{\pgfqpoint{0.069444in}{0.018417in}}{\pgfqpoint{0.062127in}{0.036082in}}{\pgfqpoint{0.049105in}{0.049105in}}%
\pgfpathcurveto{\pgfqpoint{0.036082in}{0.062127in}}{\pgfqpoint{0.018417in}{0.069444in}}{\pgfqpoint{0.000000in}{0.069444in}}%
\pgfpathcurveto{\pgfqpoint{-0.018417in}{0.069444in}}{\pgfqpoint{-0.036082in}{0.062127in}}{\pgfqpoint{-0.049105in}{0.049105in}}%
\pgfpathcurveto{\pgfqpoint{-0.062127in}{0.036082in}}{\pgfqpoint{-0.069444in}{0.018417in}}{\pgfqpoint{-0.069444in}{0.000000in}}%
\pgfpathcurveto{\pgfqpoint{-0.069444in}{-0.018417in}}{\pgfqpoint{-0.062127in}{-0.036082in}}{\pgfqpoint{-0.049105in}{-0.049105in}}%
\pgfpathcurveto{\pgfqpoint{-0.036082in}{-0.062127in}}{\pgfqpoint{-0.018417in}{-0.069444in}}{\pgfqpoint{0.000000in}{-0.069444in}}%
\pgfpathclose%
\pgfusepath{fill}%
}%
\begin{pgfscope}%
\pgfsys@transformshift{4.643870in}{2.120442in}%
\pgfsys@useobject{currentmarker}{}%
\end{pgfscope}%
\begin{pgfscope}%
\pgfsys@transformshift{4.583697in}{2.234294in}%
\pgfsys@useobject{currentmarker}{}%
\end{pgfscope}%
\begin{pgfscope}%
\pgfsys@transformshift{0.696275in}{4.475852in}%
\pgfsys@useobject{currentmarker}{}%
\end{pgfscope}%
\begin{pgfscope}%
\pgfsys@transformshift{0.730036in}{4.652099in}%
\pgfsys@useobject{currentmarker}{}%
\end{pgfscope}%
\begin{pgfscope}%
\pgfsys@transformshift{1.145169in}{1.372517in}%
\pgfsys@useobject{currentmarker}{}%
\end{pgfscope}%
\begin{pgfscope}%
\pgfsys@transformshift{1.194110in}{0.990233in}%
\pgfsys@useobject{currentmarker}{}%
\end{pgfscope}%
\end{pgfscope}%
\begin{pgfscope}%
\pgfsetrectcap%
\pgfsetmiterjoin%
\pgfsetlinewidth{1.003750pt}%
\definecolor{currentstroke}{rgb}{0.800000,0.800000,0.800000}%
\pgfsetstrokecolor{currentstroke}%
\pgfsetdash{}{0pt}%
\pgfpathmoveto{\pgfqpoint{0.344444in}{0.344444in}}%
\pgfpathlineto{\pgfqpoint{0.344444in}{4.850000in}}%
\pgfusepath{stroke}%
\end{pgfscope}%
\begin{pgfscope}%
\pgfsetrectcap%
\pgfsetmiterjoin%
\pgfsetlinewidth{1.003750pt}%
\definecolor{currentstroke}{rgb}{0.800000,0.800000,0.800000}%
\pgfsetstrokecolor{currentstroke}%
\pgfsetdash{}{0pt}%
\pgfpathmoveto{\pgfqpoint{0.344444in}{0.344444in}}%
\pgfpathlineto{\pgfqpoint{4.850000in}{0.344444in}}%
\pgfusepath{stroke}%
\end{pgfscope}%
\end{pgfpicture}%
\makeatother%
\endgroup%
};
		
		\node (cbar) at (14.5,-8) {
\begingroup%
\makeatletter%
\begin{pgfpicture}%
\pgfpathrectangle{\pgfpointorigin}{\pgfqpoint{0.650229in}{1.480710in}}%
\pgfusepath{use as bounding box, clip}%
\begin{pgfscope}%
\pgfsetbuttcap%
\pgfsetmiterjoin%
\pgfsetlinewidth{0.000000pt}%
\definecolor{currentstroke}{rgb}{0.000000,0.000000,0.000000}%
\pgfsetstrokecolor{currentstroke}%
\pgfsetstrokeopacity{0.000000}%
\pgfsetdash{}{0pt}%
\pgfpathmoveto{\pgfqpoint{0.000000in}{0.000000in}}%
\pgfpathlineto{\pgfqpoint{0.650229in}{0.000000in}}%
\pgfpathlineto{\pgfqpoint{0.650229in}{1.480710in}}%
\pgfpathlineto{\pgfqpoint{0.000000in}{1.480710in}}%
\pgfpathclose%
\pgfusepath{}%
\end{pgfscope}%
\begin{pgfscope}%
\pgfpathrectangle{\pgfqpoint{0.100000in}{0.171373in}}{\pgfqpoint{0.093750in}{1.161111in}} %
\pgfusepath{clip}%
\pgfsetbuttcap%
\pgfsetmiterjoin%
\definecolor{currentfill}{rgb}{1.000000,1.000000,1.000000}%
\pgfsetfillcolor{currentfill}%
\pgfsetlinewidth{0.010037pt}%
\definecolor{currentstroke}{rgb}{1.000000,1.000000,1.000000}%
\pgfsetstrokecolor{currentstroke}%
\pgfsetdash{}{0pt}%
\pgfpathmoveto{\pgfqpoint{0.100000in}{0.171373in}}%
\pgfpathlineto{\pgfqpoint{0.100000in}{0.175909in}}%
\pgfpathlineto{\pgfqpoint{0.100000in}{1.327949in}}%
\pgfpathlineto{\pgfqpoint{0.100000in}{1.332484in}}%
\pgfpathlineto{\pgfqpoint{0.193750in}{1.332484in}}%
\pgfpathlineto{\pgfqpoint{0.193750in}{1.327949in}}%
\pgfpathlineto{\pgfqpoint{0.193750in}{0.175909in}}%
\pgfpathlineto{\pgfqpoint{0.193750in}{0.171373in}}%
\pgfpathclose%
\pgfusepath{stroke,fill}%
\end{pgfscope}%
\begin{pgfscope}%
\definecolor{textcolor}{rgb}{0.150000,0.150000,0.150000}%
\pgfsetstrokecolor{textcolor}%
\pgfsetfillcolor{textcolor}%
\pgftext[x=0.290972in,y=0.123148in,left,base]{\color{textcolor}\sffamily\fontsize{10.000000}{12.000000}\selectfont 0.00}%
\end{pgfscope}%
\begin{pgfscope}%
\definecolor{textcolor}{rgb}{0.150000,0.150000,0.150000}%
\pgfsetstrokecolor{textcolor}%
\pgfsetfillcolor{textcolor}%
\pgftext[x=0.290972in,y=0.413426in,left,base]{\color{textcolor}\sffamily\fontsize{10.000000}{12.000000}\selectfont 0.25}%
\end{pgfscope}%
\begin{pgfscope}%
\definecolor{textcolor}{rgb}{0.150000,0.150000,0.150000}%
\pgfsetstrokecolor{textcolor}%
\pgfsetfillcolor{textcolor}%
\pgftext[x=0.290972in,y=0.703704in,left,base]{\color{textcolor}\sffamily\fontsize{10.000000}{12.000000}\selectfont 0.50}%
\end{pgfscope}%
\begin{pgfscope}%
\definecolor{textcolor}{rgb}{0.150000,0.150000,0.150000}%
\pgfsetstrokecolor{textcolor}%
\pgfsetfillcolor{textcolor}%
\pgftext[x=0.290972in,y=0.993981in,left,base]{\color{textcolor}\sffamily\fontsize{10.000000}{12.000000}\selectfont 0.75}%
\end{pgfscope}%
\begin{pgfscope}%
\definecolor{textcolor}{rgb}{0.150000,0.150000,0.150000}%
\pgfsetstrokecolor{textcolor}%
\pgfsetfillcolor{textcolor}%
\pgftext[x=0.290972in,y=1.284259in,left,base]{\color{textcolor}\sffamily\fontsize{10.000000}{12.000000}\selectfont 1.00}%
\end{pgfscope}%
\begin{pgfscope}%
\pgfsys@transformshift{0.097222in}{0.175154in}%
\pgftext[left,bottom]{\pgfimage[interpolate=true,width=0.097222in,height=1.166667in]{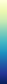}}%
\end{pgfscope}%
\begin{pgfscope}%
\pgfsetbuttcap%
\pgfsetmiterjoin%
\pgfsetlinewidth{1.003750pt}%
\definecolor{currentstroke}{rgb}{0.800000,0.800000,0.800000}%
\pgfsetstrokecolor{currentstroke}%
\pgfsetdash{}{0pt}%
\pgfpathmoveto{\pgfqpoint{0.100000in}{0.171373in}}%
\pgfpathlineto{\pgfqpoint{0.100000in}{0.175909in}}%
\pgfpathlineto{\pgfqpoint{0.100000in}{1.327949in}}%
\pgfpathlineto{\pgfqpoint{0.100000in}{1.332484in}}%
\pgfpathlineto{\pgfqpoint{0.193750in}{1.332484in}}%
\pgfpathlineto{\pgfqpoint{0.193750in}{1.327949in}}%
\pgfpathlineto{\pgfqpoint{0.193750in}{0.175909in}}%
\pgfpathlineto{\pgfqpoint{0.193750in}{0.171373in}}%
\pgfpathclose%
\pgfusepath{stroke}%
\end{pgfscope}%
\end{pgfpicture}%
\makeatother%
\endgroup%
};
		
		\node (corrspace0) at (10.2,-0.9) {};
		\node (corrspace1) at (10.5,-2.2) {};
		\node (corrspace2) at (-8.9,9.4) {};
		\node (corrspace3) at (-8.4,10.8) {};
		\node (corrspace4) at (-6.7,-5.1) {};
		\node (corrspace5) at (-6.,-7.6) {};

		\node (pixel0) at (8.5,3) {
\begingroup%
\makeatletter%
\begin{pgfpicture}%
\pgfpathrectangle{\pgfpointorigin}{\pgfqpoint{0.905556in}{1.090162in}}%
\pgfusepath{use as bounding box, clip}%
\begin{pgfscope}%
\pgfsetbuttcap%
\pgfsetmiterjoin%
\pgfsetlinewidth{0.000000pt}%
\definecolor{currentstroke}{rgb}{0.000000,0.000000,0.000000}%
\pgfsetstrokecolor{currentstroke}%
\pgfsetstrokeopacity{0.000000}%
\pgfsetdash{}{0pt}%
\pgfpathmoveto{\pgfqpoint{-0.000000in}{0.000000in}}%
\pgfpathlineto{\pgfqpoint{0.905556in}{0.000000in}}%
\pgfpathlineto{\pgfqpoint{0.905556in}{1.090162in}}%
\pgfpathlineto{\pgfqpoint{-0.000000in}{1.090162in}}%
\pgfpathclose%
\pgfusepath{}%
\end{pgfscope}%
\begin{pgfscope}%
\pgfsetbuttcap%
\pgfsetmiterjoin%
\pgfsetlinewidth{0.000000pt}%
\definecolor{currentstroke}{rgb}{0.000000,0.000000,0.000000}%
\pgfsetstrokecolor{currentstroke}%
\pgfsetstrokeopacity{0.000000}%
\pgfsetdash{}{0pt}%
\pgfpathmoveto{\pgfqpoint{0.100000in}{0.100000in}}%
\pgfpathlineto{\pgfqpoint{0.805556in}{0.100000in}}%
\pgfpathlineto{\pgfqpoint{0.805556in}{0.805556in}}%
\pgfpathlineto{\pgfqpoint{0.100000in}{0.805556in}}%
\pgfpathclose%
\pgfusepath{}%
\end{pgfscope}%
\begin{pgfscope}%
\pgfpathrectangle{\pgfqpoint{0.100000in}{0.100000in}}{\pgfqpoint{0.705556in}{0.705556in}} %
\pgfusepath{clip}%
\pgfsys@transformshift{0.100000in}{0.100000in}%
\pgftext[left,bottom]{\pgfimage[interpolate=true,width=0.708333in,height=0.708333in]{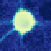}}%
\end{pgfscope}%
\begin{pgfscope}%
\pgfpathrectangle{\pgfqpoint{0.100000in}{0.100000in}}{\pgfqpoint{0.705556in}{0.705556in}} %
\pgfusepath{clip}%
\pgfsetroundcap%
\pgfsetroundjoin%
\pgfsetlinewidth{1.756562pt}%
\definecolor{currentstroke}{rgb}{1.000000,0.000000,0.000000}%
\pgfsetstrokecolor{currentstroke}%
\pgfsetdash{}{0pt}%
\pgfpathmoveto{\pgfqpoint{0.430018in}{0.634857in}}%
\pgfusepath{stroke}%
\end{pgfscope}%
\begin{pgfscope}%
\pgfpathrectangle{\pgfqpoint{0.100000in}{0.100000in}}{\pgfqpoint{0.705556in}{0.705556in}} %
\pgfusepath{clip}%
\pgfsetbuttcap%
\pgfsetmiterjoin%
\definecolor{currentfill}{rgb}{1.000000,0.000000,0.000000}%
\pgfsetfillcolor{currentfill}%
\pgfsetlinewidth{0.000000pt}%
\definecolor{currentstroke}{rgb}{1.000000,0.000000,0.000000}%
\pgfsetstrokecolor{currentstroke}%
\pgfsetdash{}{0pt}%
\pgfsys@defobject{currentmarker}{\pgfqpoint{-0.034722in}{-0.034722in}}{\pgfqpoint{0.034722in}{0.034722in}}{%
\pgfpathmoveto{\pgfqpoint{-0.034722in}{-0.034722in}}%
\pgfpathlineto{\pgfqpoint{0.034722in}{-0.034722in}}%
\pgfpathlineto{\pgfqpoint{0.034722in}{0.034722in}}%
\pgfpathlineto{\pgfqpoint{-0.034722in}{0.034722in}}%
\pgfpathclose%
\pgfusepath{fill}%
}%
\begin{pgfscope}%
\pgfsys@transformshift{0.430018in}{0.634857in}%
\pgfsys@useobject{currentmarker}{}%
\end{pgfscope}%
\end{pgfscope}%
\begin{pgfscope}%
\definecolor{textcolor}{rgb}{0.150000,0.150000,0.150000}%
\pgfsetstrokecolor{textcolor}%
\pgfsetfillcolor{textcolor}%
\pgftext[x=0.452778in,y=0.888889in,,base]{\color{textcolor}\sffamily\fontsize{10.500000}{12.600000}\selectfont Pixel 1}%
\end{pgfscope}%
\end{pgfpicture}%
\makeatother%
\endgroup%
};
		\node (pixel1) at (8.5,-6) {
\begingroup%
\makeatletter%
\begin{pgfpicture}%
\pgfpathrectangle{\pgfpointorigin}{\pgfqpoint{0.905556in}{1.090162in}}%
\pgfusepath{use as bounding box, clip}%
\begin{pgfscope}%
\pgfsetbuttcap%
\pgfsetmiterjoin%
\pgfsetlinewidth{0.000000pt}%
\definecolor{currentstroke}{rgb}{0.000000,0.000000,0.000000}%
\pgfsetstrokecolor{currentstroke}%
\pgfsetstrokeopacity{0.000000}%
\pgfsetdash{}{0pt}%
\pgfpathmoveto{\pgfqpoint{-0.000000in}{0.000000in}}%
\pgfpathlineto{\pgfqpoint{0.905556in}{0.000000in}}%
\pgfpathlineto{\pgfqpoint{0.905556in}{1.090162in}}%
\pgfpathlineto{\pgfqpoint{-0.000000in}{1.090162in}}%
\pgfpathclose%
\pgfusepath{}%
\end{pgfscope}%
\begin{pgfscope}%
\pgfsetbuttcap%
\pgfsetmiterjoin%
\pgfsetlinewidth{0.000000pt}%
\definecolor{currentstroke}{rgb}{0.000000,0.000000,0.000000}%
\pgfsetstrokecolor{currentstroke}%
\pgfsetstrokeopacity{0.000000}%
\pgfsetdash{}{0pt}%
\pgfpathmoveto{\pgfqpoint{0.100000in}{0.100000in}}%
\pgfpathlineto{\pgfqpoint{0.805556in}{0.100000in}}%
\pgfpathlineto{\pgfqpoint{0.805556in}{0.805556in}}%
\pgfpathlineto{\pgfqpoint{0.100000in}{0.805556in}}%
\pgfpathclose%
\pgfusepath{}%
\end{pgfscope}%
\begin{pgfscope}%
\pgfpathrectangle{\pgfqpoint{0.100000in}{0.100000in}}{\pgfqpoint{0.705556in}{0.705556in}} %
\pgfusepath{clip}%
\pgfsys@transformshift{0.100000in}{0.100000in}%
\pgftext[left,bottom]{\pgfimage[interpolate=true,width=0.708333in,height=0.708333in]{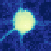}}%
\end{pgfscope}%
\begin{pgfscope}%
\pgfpathrectangle{\pgfqpoint{0.100000in}{0.100000in}}{\pgfqpoint{0.705556in}{0.705556in}} %
\pgfusepath{clip}%
\pgfsetroundcap%
\pgfsetroundjoin%
\pgfsetlinewidth{1.756562pt}%
\definecolor{currentstroke}{rgb}{1.000000,0.000000,0.000000}%
\pgfsetstrokecolor{currentstroke}%
\pgfsetdash{}{0pt}%
\pgfpathmoveto{\pgfqpoint{0.452778in}{0.452778in}}%
\pgfusepath{stroke}%
\end{pgfscope}%
\begin{pgfscope}%
\pgfpathrectangle{\pgfqpoint{0.100000in}{0.100000in}}{\pgfqpoint{0.705556in}{0.705556in}} %
\pgfusepath{clip}%
\pgfsetbuttcap%
\pgfsetmiterjoin%
\definecolor{currentfill}{rgb}{1.000000,0.000000,0.000000}%
\pgfsetfillcolor{currentfill}%
\pgfsetlinewidth{0.000000pt}%
\definecolor{currentstroke}{rgb}{1.000000,0.000000,0.000000}%
\pgfsetstrokecolor{currentstroke}%
\pgfsetdash{}{0pt}%
\pgfsys@defobject{currentmarker}{\pgfqpoint{-0.034722in}{-0.034722in}}{\pgfqpoint{0.034722in}{0.034722in}}{%
\pgfpathmoveto{\pgfqpoint{-0.034722in}{-0.034722in}}%
\pgfpathlineto{\pgfqpoint{0.034722in}{-0.034722in}}%
\pgfpathlineto{\pgfqpoint{0.034722in}{0.034722in}}%
\pgfpathlineto{\pgfqpoint{-0.034722in}{0.034722in}}%
\pgfpathclose%
\pgfusepath{fill}%
}%
\begin{pgfscope}%
\pgfsys@transformshift{0.452778in}{0.452778in}%
\pgfsys@useobject{currentmarker}{}%
\end{pgfscope}%
\end{pgfscope}%
\begin{pgfscope}%
\definecolor{textcolor}{rgb}{0.150000,0.150000,0.150000}%
\pgfsetstrokecolor{textcolor}%
\pgfsetfillcolor{textcolor}%
\pgftext[x=0.452778in,y=0.888889in,,base]{\color{textcolor}\sffamily\fontsize{10.500000}{12.600000}\selectfont Pixel 2}%
\end{pgfscope}%
\end{pgfpicture}%
\makeatother%
\endgroup%
};
		\node (pixel2) at (-7.5,5) {
\begingroup%
\makeatletter%
\begin{pgfpicture}%
\pgfpathrectangle{\pgfpointorigin}{\pgfqpoint{0.905556in}{1.090162in}}%
\pgfusepath{use as bounding box, clip}%
\begin{pgfscope}%
\pgfsetbuttcap%
\pgfsetmiterjoin%
\pgfsetlinewidth{0.000000pt}%
\definecolor{currentstroke}{rgb}{0.000000,0.000000,0.000000}%
\pgfsetstrokecolor{currentstroke}%
\pgfsetstrokeopacity{0.000000}%
\pgfsetdash{}{0pt}%
\pgfpathmoveto{\pgfqpoint{-0.000000in}{0.000000in}}%
\pgfpathlineto{\pgfqpoint{0.905556in}{0.000000in}}%
\pgfpathlineto{\pgfqpoint{0.905556in}{1.090162in}}%
\pgfpathlineto{\pgfqpoint{-0.000000in}{1.090162in}}%
\pgfpathclose%
\pgfusepath{}%
\end{pgfscope}%
\begin{pgfscope}%
\pgfsetbuttcap%
\pgfsetmiterjoin%
\pgfsetlinewidth{0.000000pt}%
\definecolor{currentstroke}{rgb}{0.000000,0.000000,0.000000}%
\pgfsetstrokecolor{currentstroke}%
\pgfsetstrokeopacity{0.000000}%
\pgfsetdash{}{0pt}%
\pgfpathmoveto{\pgfqpoint{0.100000in}{0.100000in}}%
\pgfpathlineto{\pgfqpoint{0.805556in}{0.100000in}}%
\pgfpathlineto{\pgfqpoint{0.805556in}{0.805556in}}%
\pgfpathlineto{\pgfqpoint{0.100000in}{0.805556in}}%
\pgfpathclose%
\pgfusepath{}%
\end{pgfscope}%
\begin{pgfscope}%
\pgfpathrectangle{\pgfqpoint{0.100000in}{0.100000in}}{\pgfqpoint{0.705556in}{0.705556in}} %
\pgfusepath{clip}%
\pgfsys@transformshift{0.100000in}{0.100000in}%
\pgftext[left,bottom]{\pgfimage[interpolate=true,width=0.708333in,height=0.708333in]{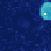}}%
\end{pgfscope}%
\begin{pgfscope}%
\pgfpathrectangle{\pgfqpoint{0.100000in}{0.100000in}}{\pgfqpoint{0.705556in}{0.705556in}} %
\pgfusepath{clip}%
\pgfsetroundcap%
\pgfsetroundjoin%
\pgfsetlinewidth{1.756562pt}%
\definecolor{currentstroke}{rgb}{1.000000,0.000000,0.000000}%
\pgfsetstrokecolor{currentstroke}%
\pgfsetdash{}{0pt}%
\pgfpathmoveto{\pgfqpoint{0.725896in}{0.612097in}}%
\pgfusepath{stroke}%
\end{pgfscope}%
\begin{pgfscope}%
\pgfpathrectangle{\pgfqpoint{0.100000in}{0.100000in}}{\pgfqpoint{0.705556in}{0.705556in}} %
\pgfusepath{clip}%
\pgfsetbuttcap%
\pgfsetmiterjoin%
\definecolor{currentfill}{rgb}{1.000000,0.000000,0.000000}%
\pgfsetfillcolor{currentfill}%
\pgfsetlinewidth{0.000000pt}%
\definecolor{currentstroke}{rgb}{1.000000,0.000000,0.000000}%
\pgfsetstrokecolor{currentstroke}%
\pgfsetdash{}{0pt}%
\pgfsys@defobject{currentmarker}{\pgfqpoint{-0.034722in}{-0.034722in}}{\pgfqpoint{0.034722in}{0.034722in}}{%
\pgfpathmoveto{\pgfqpoint{-0.034722in}{-0.034722in}}%
\pgfpathlineto{\pgfqpoint{0.034722in}{-0.034722in}}%
\pgfpathlineto{\pgfqpoint{0.034722in}{0.034722in}}%
\pgfpathlineto{\pgfqpoint{-0.034722in}{0.034722in}}%
\pgfpathclose%
\pgfusepath{fill}%
}%
\begin{pgfscope}%
\pgfsys@transformshift{0.725896in}{0.612097in}%
\pgfsys@useobject{currentmarker}{}%
\end{pgfscope}%
\end{pgfscope}%
\begin{pgfscope}%
\definecolor{textcolor}{rgb}{0.150000,0.150000,0.150000}%
\pgfsetstrokecolor{textcolor}%
\pgfsetfillcolor{textcolor}%
\pgftext[x=0.452778in,y=0.888889in,,base]{\color{textcolor}\sffamily\fontsize{10.500000}{12.600000}\selectfont Pixel 3}%
\end{pgfscope}%
\end{pgfpicture}%
\makeatother%
\endgroup%
};
		\node (pixel3) at (-2,10) {
\begingroup%
\makeatletter%
\begin{pgfpicture}%
\pgfpathrectangle{\pgfpointorigin}{\pgfqpoint{0.905556in}{1.090162in}}%
\pgfusepath{use as bounding box, clip}%
\begin{pgfscope}%
\pgfsetbuttcap%
\pgfsetmiterjoin%
\pgfsetlinewidth{0.000000pt}%
\definecolor{currentstroke}{rgb}{0.000000,0.000000,0.000000}%
\pgfsetstrokecolor{currentstroke}%
\pgfsetstrokeopacity{0.000000}%
\pgfsetdash{}{0pt}%
\pgfpathmoveto{\pgfqpoint{-0.000000in}{0.000000in}}%
\pgfpathlineto{\pgfqpoint{0.905556in}{0.000000in}}%
\pgfpathlineto{\pgfqpoint{0.905556in}{1.090162in}}%
\pgfpathlineto{\pgfqpoint{-0.000000in}{1.090162in}}%
\pgfpathclose%
\pgfusepath{}%
\end{pgfscope}%
\begin{pgfscope}%
\pgfsetbuttcap%
\pgfsetmiterjoin%
\pgfsetlinewidth{0.000000pt}%
\definecolor{currentstroke}{rgb}{0.000000,0.000000,0.000000}%
\pgfsetstrokecolor{currentstroke}%
\pgfsetstrokeopacity{0.000000}%
\pgfsetdash{}{0pt}%
\pgfpathmoveto{\pgfqpoint{0.100000in}{0.100000in}}%
\pgfpathlineto{\pgfqpoint{0.805556in}{0.100000in}}%
\pgfpathlineto{\pgfqpoint{0.805556in}{0.805556in}}%
\pgfpathlineto{\pgfqpoint{0.100000in}{0.805556in}}%
\pgfpathclose%
\pgfusepath{}%
\end{pgfscope}%
\begin{pgfscope}%
\pgfpathrectangle{\pgfqpoint{0.100000in}{0.100000in}}{\pgfqpoint{0.705556in}{0.705556in}} %
\pgfusepath{clip}%
\pgfsys@transformshift{0.100000in}{0.100000in}%
\pgftext[left,bottom]{\pgfimage[interpolate=true,width=0.708333in,height=0.708333in]{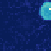}}%
\end{pgfscope}%
\begin{pgfscope}%
\pgfpathrectangle{\pgfqpoint{0.100000in}{0.100000in}}{\pgfqpoint{0.705556in}{0.705556in}} %
\pgfusepath{clip}%
\pgfsetroundcap%
\pgfsetroundjoin%
\pgfsetlinewidth{1.756562pt}%
\definecolor{currentstroke}{rgb}{1.000000,0.000000,0.000000}%
\pgfsetstrokecolor{currentstroke}%
\pgfsetdash{}{0pt}%
\pgfpathmoveto{\pgfqpoint{0.794176in}{0.680376in}}%
\pgfusepath{stroke}%
\end{pgfscope}%
\begin{pgfscope}%
\pgfpathrectangle{\pgfqpoint{0.100000in}{0.100000in}}{\pgfqpoint{0.705556in}{0.705556in}} %
\pgfusepath{clip}%
\pgfsetbuttcap%
\pgfsetmiterjoin%
\definecolor{currentfill}{rgb}{1.000000,0.000000,0.000000}%
\pgfsetfillcolor{currentfill}%
\pgfsetlinewidth{0.000000pt}%
\definecolor{currentstroke}{rgb}{1.000000,0.000000,0.000000}%
\pgfsetstrokecolor{currentstroke}%
\pgfsetdash{}{0pt}%
\pgfsys@defobject{currentmarker}{\pgfqpoint{-0.034722in}{-0.034722in}}{\pgfqpoint{0.034722in}{0.034722in}}{%
\pgfpathmoveto{\pgfqpoint{-0.034722in}{-0.034722in}}%
\pgfpathlineto{\pgfqpoint{0.034722in}{-0.034722in}}%
\pgfpathlineto{\pgfqpoint{0.034722in}{0.034722in}}%
\pgfpathlineto{\pgfqpoint{-0.034722in}{0.034722in}}%
\pgfpathclose%
\pgfusepath{fill}%
}%
\begin{pgfscope}%
\pgfsys@transformshift{0.794176in}{0.680376in}%
\pgfsys@useobject{currentmarker}{}%
\end{pgfscope}%
\end{pgfscope}%
\begin{pgfscope}%
\definecolor{textcolor}{rgb}{0.150000,0.150000,0.150000}%
\pgfsetstrokecolor{textcolor}%
\pgfsetfillcolor{textcolor}%
\pgftext[x=0.452778in,y=0.888889in,,base]{\color{textcolor}\sffamily\fontsize{10.500000}{12.600000}\selectfont Pixel 4}%
\end{pgfscope}%
\end{pgfpicture}%
\makeatother%
\endgroup%
};
		\node (pixel4) at (-4.5,-1) {
\begingroup%
\makeatletter%
\begin{pgfpicture}%
\pgfpathrectangle{\pgfpointorigin}{\pgfqpoint{0.905556in}{1.090162in}}%
\pgfusepath{use as bounding box, clip}%
\begin{pgfscope}%
\pgfsetbuttcap%
\pgfsetmiterjoin%
\pgfsetlinewidth{0.000000pt}%
\definecolor{currentstroke}{rgb}{0.000000,0.000000,0.000000}%
\pgfsetstrokecolor{currentstroke}%
\pgfsetstrokeopacity{0.000000}%
\pgfsetdash{}{0pt}%
\pgfpathmoveto{\pgfqpoint{-0.000000in}{0.000000in}}%
\pgfpathlineto{\pgfqpoint{0.905556in}{0.000000in}}%
\pgfpathlineto{\pgfqpoint{0.905556in}{1.090162in}}%
\pgfpathlineto{\pgfqpoint{-0.000000in}{1.090162in}}%
\pgfpathclose%
\pgfusepath{}%
\end{pgfscope}%
\begin{pgfscope}%
\pgfsetbuttcap%
\pgfsetmiterjoin%
\pgfsetlinewidth{0.000000pt}%
\definecolor{currentstroke}{rgb}{0.000000,0.000000,0.000000}%
\pgfsetstrokecolor{currentstroke}%
\pgfsetstrokeopacity{0.000000}%
\pgfsetdash{}{0pt}%
\pgfpathmoveto{\pgfqpoint{0.100000in}{0.100000in}}%
\pgfpathlineto{\pgfqpoint{0.805556in}{0.100000in}}%
\pgfpathlineto{\pgfqpoint{0.805556in}{0.805556in}}%
\pgfpathlineto{\pgfqpoint{0.100000in}{0.805556in}}%
\pgfpathclose%
\pgfusepath{}%
\end{pgfscope}%
\begin{pgfscope}%
\pgfpathrectangle{\pgfqpoint{0.100000in}{0.100000in}}{\pgfqpoint{0.705556in}{0.705556in}} %
\pgfusepath{clip}%
\pgfsys@transformshift{0.100000in}{0.100000in}%
\pgftext[left,bottom]{\pgfimage[interpolate=true,width=0.708333in,height=0.708333in]{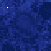}}%
\end{pgfscope}%
\begin{pgfscope}%
\pgfpathrectangle{\pgfqpoint{0.100000in}{0.100000in}}{\pgfqpoint{0.705556in}{0.705556in}} %
\pgfusepath{clip}%
\pgfsetroundcap%
\pgfsetroundjoin%
\pgfsetlinewidth{1.756562pt}%
\definecolor{currentstroke}{rgb}{1.000000,0.000000,0.000000}%
\pgfsetstrokecolor{currentstroke}%
\pgfsetdash{}{0pt}%
\pgfpathmoveto{\pgfqpoint{0.202419in}{0.703136in}}%
\pgfusepath{stroke}%
\end{pgfscope}%
\begin{pgfscope}%
\pgfpathrectangle{\pgfqpoint{0.100000in}{0.100000in}}{\pgfqpoint{0.705556in}{0.705556in}} %
\pgfusepath{clip}%
\pgfsetbuttcap%
\pgfsetmiterjoin%
\definecolor{currentfill}{rgb}{1.000000,0.000000,0.000000}%
\pgfsetfillcolor{currentfill}%
\pgfsetlinewidth{0.000000pt}%
\definecolor{currentstroke}{rgb}{1.000000,0.000000,0.000000}%
\pgfsetstrokecolor{currentstroke}%
\pgfsetdash{}{0pt}%
\pgfsys@defobject{currentmarker}{\pgfqpoint{-0.034722in}{-0.034722in}}{\pgfqpoint{0.034722in}{0.034722in}}{%
\pgfpathmoveto{\pgfqpoint{-0.034722in}{-0.034722in}}%
\pgfpathlineto{\pgfqpoint{0.034722in}{-0.034722in}}%
\pgfpathlineto{\pgfqpoint{0.034722in}{0.034722in}}%
\pgfpathlineto{\pgfqpoint{-0.034722in}{0.034722in}}%
\pgfpathclose%
\pgfusepath{fill}%
}%
\begin{pgfscope}%
\pgfsys@transformshift{0.202419in}{0.703136in}%
\pgfsys@useobject{currentmarker}{}%
\end{pgfscope}%
\end{pgfscope}%
\begin{pgfscope}%
\definecolor{textcolor}{rgb}{0.150000,0.150000,0.150000}%
\pgfsetstrokecolor{textcolor}%
\pgfsetfillcolor{textcolor}%
\pgftext[x=0.452778in,y=0.888889in,,base]{\color{textcolor}\sffamily\fontsize{10.500000}{12.600000}\selectfont Pixel 5}%
\end{pgfscope}%
\end{pgfpicture}%
\makeatother%
\endgroup%
};
		\node (pixel5) at (-2,-8) {
\begingroup%
\makeatletter%
\begin{pgfpicture}%
\pgfpathrectangle{\pgfpointorigin}{\pgfqpoint{0.905556in}{1.090162in}}%
\pgfusepath{use as bounding box, clip}%
\begin{pgfscope}%
\pgfsetbuttcap%
\pgfsetmiterjoin%
\pgfsetlinewidth{0.000000pt}%
\definecolor{currentstroke}{rgb}{0.000000,0.000000,0.000000}%
\pgfsetstrokecolor{currentstroke}%
\pgfsetstrokeopacity{0.000000}%
\pgfsetdash{}{0pt}%
\pgfpathmoveto{\pgfqpoint{-0.000000in}{0.000000in}}%
\pgfpathlineto{\pgfqpoint{0.905556in}{0.000000in}}%
\pgfpathlineto{\pgfqpoint{0.905556in}{1.090162in}}%
\pgfpathlineto{\pgfqpoint{-0.000000in}{1.090162in}}%
\pgfpathclose%
\pgfusepath{}%
\end{pgfscope}%
\begin{pgfscope}%
\pgfsetbuttcap%
\pgfsetmiterjoin%
\pgfsetlinewidth{0.000000pt}%
\definecolor{currentstroke}{rgb}{0.000000,0.000000,0.000000}%
\pgfsetstrokecolor{currentstroke}%
\pgfsetstrokeopacity{0.000000}%
\pgfsetdash{}{0pt}%
\pgfpathmoveto{\pgfqpoint{0.100000in}{0.100000in}}%
\pgfpathlineto{\pgfqpoint{0.805556in}{0.100000in}}%
\pgfpathlineto{\pgfqpoint{0.805556in}{0.805556in}}%
\pgfpathlineto{\pgfqpoint{0.100000in}{0.805556in}}%
\pgfpathclose%
\pgfusepath{}%
\end{pgfscope}%
\begin{pgfscope}%
\pgfpathrectangle{\pgfqpoint{0.100000in}{0.100000in}}{\pgfqpoint{0.705556in}{0.705556in}} %
\pgfusepath{clip}%
\pgfsys@transformshift{0.100000in}{0.100000in}%
\pgftext[left,bottom]{\pgfimage[interpolate=true,width=0.708333in,height=0.708333in]{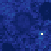}}%
\end{pgfscope}%
\begin{pgfscope}%
\pgfpathrectangle{\pgfqpoint{0.100000in}{0.100000in}}{\pgfqpoint{0.705556in}{0.705556in}} %
\pgfusepath{clip}%
\pgfsetroundcap%
\pgfsetroundjoin%
\pgfsetlinewidth{1.756562pt}%
\definecolor{currentstroke}{rgb}{1.000000,0.000000,0.000000}%
\pgfsetstrokecolor{currentstroke}%
\pgfsetdash{}{0pt}%
\pgfpathmoveto{\pgfqpoint{0.680376in}{0.338978in}}%
\pgfusepath{stroke}%
\end{pgfscope}%
\begin{pgfscope}%
\pgfpathrectangle{\pgfqpoint{0.100000in}{0.100000in}}{\pgfqpoint{0.705556in}{0.705556in}} %
\pgfusepath{clip}%
\pgfsetbuttcap%
\pgfsetmiterjoin%
\definecolor{currentfill}{rgb}{1.000000,0.000000,0.000000}%
\pgfsetfillcolor{currentfill}%
\pgfsetlinewidth{0.000000pt}%
\definecolor{currentstroke}{rgb}{1.000000,0.000000,0.000000}%
\pgfsetstrokecolor{currentstroke}%
\pgfsetdash{}{0pt}%
\pgfsys@defobject{currentmarker}{\pgfqpoint{-0.034722in}{-0.034722in}}{\pgfqpoint{0.034722in}{0.034722in}}{%
\pgfpathmoveto{\pgfqpoint{-0.034722in}{-0.034722in}}%
\pgfpathlineto{\pgfqpoint{0.034722in}{-0.034722in}}%
\pgfpathlineto{\pgfqpoint{0.034722in}{0.034722in}}%
\pgfpathlineto{\pgfqpoint{-0.034722in}{0.034722in}}%
\pgfpathclose%
\pgfusepath{fill}%
}%
\begin{pgfscope}%
\pgfsys@transformshift{0.680376in}{0.338978in}%
\pgfsys@useobject{currentmarker}{}%
\end{pgfscope}%
\end{pgfscope}%
\begin{pgfscope}%
\definecolor{textcolor}{rgb}{0.150000,0.150000,0.150000}%
\pgfsetstrokecolor{textcolor}%
\pgfsetfillcolor{textcolor}%
\pgftext[x=0.452778in,y=0.888889in,,base]{\color{textcolor}\sffamily\fontsize{10.500000}{12.600000}\selectfont Pixel 6}%
\end{pgfscope}%
\end{pgfpicture}%
\makeatother%
\endgroup%
};
		
		\draw[dashed] (10.3,0.5) -- (corrspace0.center);
		\draw[dashed] (6.7,0.5) -- (corrspace0.center);
		
		\draw[dashed] (10.3,-3.5) -- (corrspace1.center);
		\draw[dashed] (6.7,-3.5) -- (corrspace1.center);
		
		\draw[dashed] (-5.7,7) -- (corrspace2.center);
		\draw[dashed] (-9.3,7) -- (corrspace2.center);
		
		\draw[dashed] (-4.2,11.3) -- (corrspace3.center);
		\draw[dashed] (-4.2,7.7) -- (corrspace3.center);
		
		\draw[dashed] (-6.3,-3.5) -- (corrspace4.center);
		\draw[dashed] (-2.7,-3.5) -- (corrspace4.center);
		
		\draw[dashed] (-4.1,-10.2) -- (corrspace5.center);
		\draw[dashed] (-4.2,-6.7) -- (corrspace5.center);
		\end{tikzpicture}
		\caption{Pixels (blue) plotted in a 2-dimensional PCA representation of the correlation space. Each insert visualizes the correlation vector of the associated pixel (marked in red). Pixels in the spatial footprints of each cell and the background pixels are clustered in correlation space.}
		\label{fig:corrspace}
	\end{minipage}
\end{figure}

\begin{figure}
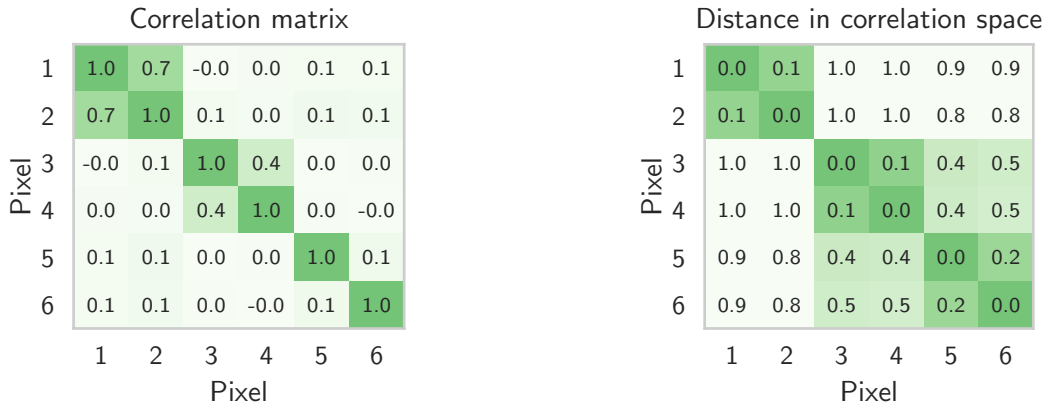

	\centering
	\begin{minipage}[t]{0.49\textwidth}
		\centering
		\input{figures/corrBetween.pgf}
		\subcaption{ High correlation values (green) indicate high similarity between the signals of a pair of pixels. Pixels that are part of the same cell have highly correlated signals. The signals of the background pixels are not correlated. See Figure \ref{fig:corrspace} for the location of the pixels. }
	\end{minipage}
	\hfill
	\begin{minipage}[t]{0.49\textwidth}
		\centering
		\input{figures/corrDistance.pgf}
		\subcaption{ Small $\ell_2$ distance values (green) indicate that pixels are close in correlation space. The distances are normalized such that the largest distance is one. Cells as well as background form clusters in correlation space. This results in a block matrix that is approximately diagonal.}
	\end{minipage}
	
	\caption{Visualization of the correlation space for a window of the 02.00 training dataset of the Neurofinder benchmark. The pixels tightly cluster into the cells and background in correlation space but not in the regular correlation matrix.}
	\label{fig:corrspaceMat}
\end{figure}

In correlation space, a pixel is represented by a vector containing the correlations between the pixel's intensity timeseries and the timeseries of {\em all pixels in the window}.
That is, every pixel $i$ is described by a vector of correlations $R_i \in [-1,1]^n$, where $n=n_1\times n_2$ is the number of pixels in the window.
The pixels that are part of a spatial footprint will have high values for the entries that correspond to pixels that are in the spatial footprint and low values for the entries of pixels that are not in the spatial footprint.
As a result, the pixels in each spatial footprint as well as the background pixels cluster in correlation space.
See Figures \ref{fig:corrspace} and \ref{fig:corrspaceMat} for visualizations of this phenomenon.

To define $w_{ij}$ we measure the Gaussian similarity between pixels $i$ and $j$ in correlation space.
Precisely, we define $w_{ij}$ as:
$$
w_{ij} := \exp\left(
- \frac{
	\left\| R_i - R_j \right\|^2_2
}{
	\sigma^2
}
\right)
$$
where $\sigma$ is a scaling parameter that is typically set to $1$.
Note that $w_{ij}$ is close to one when the distance between pixels $i$ and $j$ in correlation space is small and near zero when the distance is large.


\subsection*{Sparse Computation}
For the edge set of the graph $G=(V,E)$ on which we solve the HNC model, one option is to naively select $E$ to consist of all possible pairs of pixels.
However, this can be computationally costly and unnecessary since most pairs of pixels are highly dissimilar.
Instead, we use a methodology called \emph{sparse computation}~\cite{HocBau16,BauHocSpa16}.
Sparse computation allows us to compute only a small subset of all possible pairs.
It uses the observation that for many pairs the pairwise similarity is close to zero and that removing these edges does typically not affect the outcome.
For general machine learning, sparse computation significantly reduces the running time of similarity-based classifiers~\cite{HocBau16,BauHocSpa16}, such as SNC~\cite{YanFisHoc13}, KNN, and kernel SVM.
In sparse computation, the data, here consisting of the correlation vectors, is efficiently projected onto a low-dimensional space of dimension $p$, typically using a fast approximation of principal component analysis.
The low-dimensional space is then subdivided into $\kappa$ sections per dimension resulting in $\kappa^p$ grid blocks.
We add to the edge set $E$ edges between pixels in the same or neighboring grid blocks.

\subsection*{HNCcorr algorithm}
Having introduced the key ideas behind HNCcorr, we now describe the HNCcorr algorithm.

The HNCcorr algorithm iterates through a list of seeds.
For each seed, the algorithm performs the steps described below to identify a cell whose spatial footprint contains the seed or to conclude that no such cell exists.
The algorithm first constructs a window around the seed and picks the negative seeds.
Next, the algorithm maps each pixel to correlation space.
It then determines which edges are in the graph with the sparse computation technique.
For each of these edges, $w_{ij}$ is computed.
Now that the graph is well-defined, the algorithm calls a subroutine to solve the corresponding HNC problem for all $\lambda \ge 0$. We can obtain an optimal set $S(\lambda)$ for each value of $\lambda$, since the sets $S(\lambda)$ are nested ($\lambda _1<\lambda _2$, $S(\lambda_1) \subseteq S(\lambda _2)$) and therefore there can be at most $n$ such sets.

Finally, a simple oracle decides the best spatial footprint for the cell based on these sets $S(\lambda)$ or concludes that there is no cell located at the seed.
Currently, the oracle simply checks if the number of segmented pixels is between a given lower and upper bound based on the expected size of a cell.
In case none of the segmentations satisfy this criteria, then the oracle concludes that no cell containing the input seed exists.
In case more than one segmentation satisfies the size criteria, the oracle returns the spatial footprint where the number of segmented pixels is closest to a given expected cell size.
Note that more complex oracles can be used as well.

The output of the algorithm is either the best segmentation or an empty set indicating that no cell has been identified. This process repeats for each of the selected seeds.


\begin{figure}
	\centering
	\begin{minipage}[t]{0.48\textwidth}
		\vspace{0pt}
		\centering
\begingroup%
\makeatletter%
\begin{pgfpicture}%
\pgfpathrectangle{\pgfpointorigin}{\pgfqpoint{1.856595in}{1.447440in}}%
\pgfusepath{use as bounding box, clip}%
\begin{pgfscope}%
\pgfsetbuttcap%
\pgfsetmiterjoin%
\definecolor{currentfill}{rgb}{1.000000,1.000000,1.000000}%
\pgfsetfillcolor{currentfill}%
\pgfsetlinewidth{0.000000pt}%
\definecolor{currentstroke}{rgb}{1.000000,1.000000,1.000000}%
\pgfsetstrokecolor{currentstroke}%
\pgfsetdash{}{0pt}%
\pgfpathmoveto{\pgfqpoint{0.000000in}{0.000000in}}%
\pgfpathlineto{\pgfqpoint{1.856595in}{0.000000in}}%
\pgfpathlineto{\pgfqpoint{1.856595in}{1.447440in}}%
\pgfpathlineto{\pgfqpoint{0.000000in}{1.447440in}}%
\pgfpathclose%
\pgfusepath{fill}%
\end{pgfscope}%
\begin{pgfscope}%
\pgfsetbuttcap%
\pgfsetmiterjoin%
\definecolor{currentfill}{rgb}{1.000000,1.000000,1.000000}%
\pgfsetfillcolor{currentfill}%
\pgfsetlinewidth{0.000000pt}%
\definecolor{currentstroke}{rgb}{0.000000,0.000000,0.000000}%
\pgfsetstrokecolor{currentstroke}%
\pgfsetstrokeopacity{0.000000}%
\pgfsetdash{}{0pt}%
\pgfpathmoveto{\pgfqpoint{0.100000in}{0.171373in}}%
\pgfpathlineto{\pgfqpoint{1.227841in}{0.171373in}}%
\pgfpathlineto{\pgfqpoint{1.227841in}{1.299214in}}%
\pgfpathlineto{\pgfqpoint{0.100000in}{1.299214in}}%
\pgfpathclose%
\pgfusepath{fill}%
\end{pgfscope}%
\begin{pgfscope}%
\pgfpathrectangle{\pgfqpoint{0.100000in}{0.171373in}}{\pgfqpoint{1.127841in}{1.127841in}} %
\pgfusepath{clip}%
\pgfsys@transformshift{0.100000in}{0.171373in}%
\pgftext[left,bottom]{\pgfimage[interpolate=true,width=1.130000in,height=1.130000in]{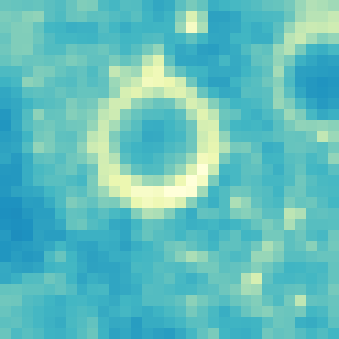}}%
\end{pgfscope}%
\begin{pgfscope}%
\pgfpathrectangle{\pgfqpoint{0.100000in}{0.171373in}}{\pgfqpoint{1.127841in}{1.127841in}} %
\pgfusepath{clip}%
\pgfsetroundcap%
\pgfsetroundjoin%
\pgfsetlinewidth{1.756562pt}%
\definecolor{currentstroke}{rgb}{1.000000,0.000000,0.000000}%
\pgfsetstrokecolor{currentstroke}%
\pgfsetdash{}{0pt}%
\pgfpathmoveto{\pgfqpoint{0.700302in}{0.698912in}}%
\pgfusepath{stroke}%
\end{pgfscope}%
\begin{pgfscope}%
\pgfpathrectangle{\pgfqpoint{0.100000in}{0.171373in}}{\pgfqpoint{1.127841in}{1.127841in}} %
\pgfusepath{clip}%
\pgfsetbuttcap%
\pgfsetmiterjoin%
\definecolor{currentfill}{rgb}{1.000000,0.000000,0.000000}%
\pgfsetfillcolor{currentfill}%
\pgfsetlinewidth{0.000000pt}%
\definecolor{currentstroke}{rgb}{1.000000,0.000000,0.000000}%
\pgfsetstrokecolor{currentstroke}%
\pgfsetdash{}{0pt}%
\pgfsys@defobject{currentmarker}{\pgfqpoint{-0.034722in}{-0.034722in}}{\pgfqpoint{0.034722in}{0.034722in}}{%
\pgfpathmoveto{\pgfqpoint{-0.034722in}{-0.034722in}}%
\pgfpathlineto{\pgfqpoint{0.034722in}{-0.034722in}}%
\pgfpathlineto{\pgfqpoint{0.034722in}{0.034722in}}%
\pgfpathlineto{\pgfqpoint{-0.034722in}{0.034722in}}%
\pgfpathclose%
\pgfusepath{fill}%
}%
\begin{pgfscope}%
\pgfsys@transformshift{0.700302in}{0.698912in}%
\pgfsys@useobject{currentmarker}{}%
\end{pgfscope}%
\end{pgfscope}%
\begin{pgfscope}%
\pgfpathrectangle{\pgfqpoint{1.327841in}{0.171373in}}{\pgfqpoint{0.112784in}{1.127841in}} %
\pgfusepath{clip}%
\pgfsetbuttcap%
\pgfsetmiterjoin%
\definecolor{currentfill}{rgb}{1.000000,1.000000,1.000000}%
\pgfsetfillcolor{currentfill}%
\pgfsetlinewidth{0.010037pt}%
\definecolor{currentstroke}{rgb}{1.000000,1.000000,1.000000}%
\pgfsetstrokecolor{currentstroke}%
\pgfsetdash{}{0pt}%
\pgfpathmoveto{\pgfqpoint{1.327841in}{0.171373in}}%
\pgfpathlineto{\pgfqpoint{1.327841in}{0.175779in}}%
\pgfpathlineto{\pgfqpoint{1.327841in}{1.294809in}}%
\pgfpathlineto{\pgfqpoint{1.327841in}{1.299214in}}%
\pgfpathlineto{\pgfqpoint{1.440625in}{1.299214in}}%
\pgfpathlineto{\pgfqpoint{1.440625in}{1.294809in}}%
\pgfpathlineto{\pgfqpoint{1.440625in}{0.175779in}}%
\pgfpathlineto{\pgfqpoint{1.440625in}{0.171373in}}%
\pgfpathclose%
\pgfusepath{stroke,fill}%
\end{pgfscope}%
\begin{pgfscope}%
\definecolor{textcolor}{rgb}{0.150000,0.150000,0.150000}%
\pgfsetstrokecolor{textcolor}%
\pgfsetfillcolor{textcolor}%
\pgftext[x=1.537847in,y=0.123148in,left,base]{\color{textcolor}\sffamily\fontsize{10.000000}{12.000000}\selectfont 0}%
\end{pgfscope}%
\begin{pgfscope}%
\definecolor{textcolor}{rgb}{0.150000,0.150000,0.150000}%
\pgfsetstrokecolor{textcolor}%
\pgfsetfillcolor{textcolor}%
\pgftext[x=1.537847in,y=0.499095in,left,base]{\color{textcolor}\sffamily\fontsize{10.000000}{12.000000}\selectfont 100}%
\end{pgfscope}%
\begin{pgfscope}%
\definecolor{textcolor}{rgb}{0.150000,0.150000,0.150000}%
\pgfsetstrokecolor{textcolor}%
\pgfsetfillcolor{textcolor}%
\pgftext[x=1.537847in,y=0.875042in,left,base]{\color{textcolor}\sffamily\fontsize{10.000000}{12.000000}\selectfont 200}%
\end{pgfscope}%
\begin{pgfscope}%
\definecolor{textcolor}{rgb}{0.150000,0.150000,0.150000}%
\pgfsetstrokecolor{textcolor}%
\pgfsetfillcolor{textcolor}%
\pgftext[x=1.537847in,y=1.250989in,left,base]{\color{textcolor}\sffamily\fontsize{10.000000}{12.000000}\selectfont 300}%
\end{pgfscope}%
\begin{pgfscope}%
\pgfsys@transformshift{1.326667in}{0.170773in}%
\pgftext[left,bottom]{\pgfimage[interpolate=true,width=0.113333in,height=1.130000in]{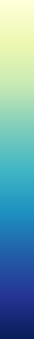}}%
\end{pgfscope}%
\begin{pgfscope}%
\pgfsetbuttcap%
\pgfsetmiterjoin%
\pgfsetlinewidth{1.003750pt}%
\definecolor{currentstroke}{rgb}{0.800000,0.800000,0.800000}%
\pgfsetstrokecolor{currentstroke}%
\pgfsetdash{}{0pt}%
\pgfpathmoveto{\pgfqpoint{1.327841in}{0.171373in}}%
\pgfpathlineto{\pgfqpoint{1.327841in}{0.175779in}}%
\pgfpathlineto{\pgfqpoint{1.327841in}{1.294809in}}%
\pgfpathlineto{\pgfqpoint{1.327841in}{1.299214in}}%
\pgfpathlineto{\pgfqpoint{1.440625in}{1.299214in}}%
\pgfpathlineto{\pgfqpoint{1.440625in}{1.294809in}}%
\pgfpathlineto{\pgfqpoint{1.440625in}{0.175779in}}%
\pgfpathlineto{\pgfqpoint{1.440625in}{0.171373in}}%
\pgfpathclose%
\pgfusepath{stroke}%
\end{pgfscope}%
\end{pgfpicture}%
\makeatother%
\endgroup%
		\subcaption{The average intensity of each pixel in the window over time. The red square marks the seed for HNCcorr.}
		\label{fig:algSeed}
	\end{minipage}
	\hfill
	\begin{minipage}[t]{0.48\textwidth}
		\vspace{0pt}
		\centering
\begingroup%
\makeatletter%
\begin{pgfpicture}%
\pgfpathrectangle{\pgfpointorigin}{\pgfqpoint{1.856595in}{1.447064in}}%
\pgfusepath{use as bounding box, clip}%
\begin{pgfscope}%
\pgfsetbuttcap%
\pgfsetmiterjoin%
\definecolor{currentfill}{rgb}{1.000000,1.000000,1.000000}%
\pgfsetfillcolor{currentfill}%
\pgfsetlinewidth{0.000000pt}%
\definecolor{currentstroke}{rgb}{1.000000,1.000000,1.000000}%
\pgfsetstrokecolor{currentstroke}%
\pgfsetdash{}{0pt}%
\pgfpathmoveto{\pgfqpoint{0.000000in}{0.000000in}}%
\pgfpathlineto{\pgfqpoint{1.856595in}{0.000000in}}%
\pgfpathlineto{\pgfqpoint{1.856595in}{1.447064in}}%
\pgfpathlineto{\pgfqpoint{0.000000in}{1.447064in}}%
\pgfpathclose%
\pgfusepath{fill}%
\end{pgfscope}%
\begin{pgfscope}%
\pgfsetbuttcap%
\pgfsetmiterjoin%
\definecolor{currentfill}{rgb}{1.000000,1.000000,1.000000}%
\pgfsetfillcolor{currentfill}%
\pgfsetlinewidth{0.000000pt}%
\definecolor{currentstroke}{rgb}{0.000000,0.000000,0.000000}%
\pgfsetstrokecolor{currentstroke}%
\pgfsetstrokeopacity{0.000000}%
\pgfsetdash{}{0pt}%
\pgfpathmoveto{\pgfqpoint{0.100000in}{0.171373in}}%
\pgfpathlineto{\pgfqpoint{1.227841in}{0.171373in}}%
\pgfpathlineto{\pgfqpoint{1.227841in}{1.299214in}}%
\pgfpathlineto{\pgfqpoint{0.100000in}{1.299214in}}%
\pgfpathclose%
\pgfusepath{fill}%
\end{pgfscope}%
\begin{pgfscope}%
\pgfpathrectangle{\pgfqpoint{0.100000in}{0.171373in}}{\pgfqpoint{1.127841in}{1.127841in}} %
\pgfusepath{clip}%
\pgfsys@transformshift{0.100000in}{0.171373in}%
\pgftext[left,bottom]{\pgfimage[interpolate=true,width=1.130000in,height=1.130000in]{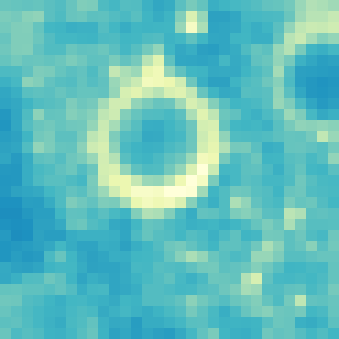}}%
\end{pgfscope}%
\begin{pgfscope}%
\pgfpathrectangle{\pgfqpoint{0.100000in}{0.171373in}}{\pgfqpoint{1.127841in}{1.127841in}} %
\pgfusepath{clip}%
\pgfsetbuttcap%
\pgfsetroundjoin%
\pgfsetlinewidth{3.011250pt}%
\definecolor{currentstroke}{rgb}{0.988235,0.552941,0.349020}%
\pgfsetstrokecolor{currentstroke}%
\pgfsetdash{}{0pt}%
\pgfpathmoveto{\pgfqpoint{0.391056in}{0.844440in}}%
\pgfpathlineto{\pgfqpoint{0.409247in}{0.826249in}}%
\pgfpathlineto{\pgfqpoint{0.427438in}{0.808058in}}%
\pgfpathlineto{\pgfqpoint{0.427438in}{0.771676in}}%
\pgfpathlineto{\pgfqpoint{0.427438in}{0.735294in}}%
\pgfpathlineto{\pgfqpoint{0.427438in}{0.698912in}}%
\pgfpathlineto{\pgfqpoint{0.445629in}{0.680721in}}%
\pgfpathlineto{\pgfqpoint{0.463820in}{0.662530in}}%
\pgfpathlineto{\pgfqpoint{0.482011in}{0.644339in}}%
\pgfpathlineto{\pgfqpoint{0.500202in}{0.626148in}}%
\pgfpathlineto{\pgfqpoint{0.518393in}{0.607957in}}%
\pgfpathlineto{\pgfqpoint{0.554775in}{0.607957in}}%
\pgfpathlineto{\pgfqpoint{0.572966in}{0.589766in}}%
\pgfpathlineto{\pgfqpoint{0.591157in}{0.571575in}}%
\pgfpathlineto{\pgfqpoint{0.627538in}{0.571575in}}%
\pgfpathlineto{\pgfqpoint{0.645729in}{0.589766in}}%
\pgfpathlineto{\pgfqpoint{0.663920in}{0.607957in}}%
\pgfpathlineto{\pgfqpoint{0.700302in}{0.607957in}}%
\pgfpathlineto{\pgfqpoint{0.718493in}{0.626148in}}%
\pgfpathlineto{\pgfqpoint{0.736684in}{0.644339in}}%
\pgfpathlineto{\pgfqpoint{0.773066in}{0.644339in}}%
\pgfpathlineto{\pgfqpoint{0.791257in}{0.662530in}}%
\pgfpathlineto{\pgfqpoint{0.791257in}{0.698912in}}%
\pgfpathlineto{\pgfqpoint{0.809448in}{0.717103in}}%
\pgfpathlineto{\pgfqpoint{0.827639in}{0.735294in}}%
\pgfpathlineto{\pgfqpoint{0.845830in}{0.753485in}}%
\pgfpathlineto{\pgfqpoint{0.864021in}{0.771676in}}%
\pgfpathlineto{\pgfqpoint{0.864021in}{0.808058in}}%
\pgfpathlineto{\pgfqpoint{0.864021in}{0.844440in}}%
\pgfpathlineto{\pgfqpoint{0.845830in}{0.862631in}}%
\pgfpathlineto{\pgfqpoint{0.827639in}{0.880822in}}%
\pgfpathlineto{\pgfqpoint{0.827639in}{0.917204in}}%
\pgfpathlineto{\pgfqpoint{0.809448in}{0.935395in}}%
\pgfpathlineto{\pgfqpoint{0.791257in}{0.953586in}}%
\pgfpathlineto{\pgfqpoint{0.773066in}{0.971777in}}%
\pgfpathlineto{\pgfqpoint{0.754875in}{0.989968in}}%
\pgfpathlineto{\pgfqpoint{0.736684in}{1.008159in}}%
\pgfpathlineto{\pgfqpoint{0.718493in}{1.026350in}}%
\pgfpathlineto{\pgfqpoint{0.700302in}{1.044540in}}%
\pgfpathlineto{\pgfqpoint{0.663920in}{1.044540in}}%
\pgfpathlineto{\pgfqpoint{0.627538in}{1.044540in}}%
\pgfpathlineto{\pgfqpoint{0.591157in}{1.044540in}}%
\pgfpathlineto{\pgfqpoint{0.554775in}{1.044540in}}%
\pgfpathlineto{\pgfqpoint{0.536584in}{1.026350in}}%
\pgfpathlineto{\pgfqpoint{0.518393in}{1.008159in}}%
\pgfpathlineto{\pgfqpoint{0.482011in}{1.008159in}}%
\pgfpathlineto{\pgfqpoint{0.463820in}{0.989968in}}%
\pgfpathlineto{\pgfqpoint{0.463820in}{0.953586in}}%
\pgfpathlineto{\pgfqpoint{0.445629in}{0.935395in}}%
\pgfpathlineto{\pgfqpoint{0.427438in}{0.917204in}}%
\pgfpathlineto{\pgfqpoint{0.427438in}{0.880822in}}%
\pgfpathlineto{\pgfqpoint{0.409247in}{0.862631in}}%
\pgfpathlineto{\pgfqpoint{0.391056in}{0.844440in}}%
\pgfusepath{stroke}%
\end{pgfscope}%
\begin{pgfscope}%
\pgfpathrectangle{\pgfqpoint{1.327841in}{0.171373in}}{\pgfqpoint{0.112784in}{1.127841in}} %
\pgfusepath{clip}%
\pgfsetbuttcap%
\pgfsetmiterjoin%
\definecolor{currentfill}{rgb}{1.000000,1.000000,1.000000}%
\pgfsetfillcolor{currentfill}%
\pgfsetlinewidth{0.010037pt}%
\definecolor{currentstroke}{rgb}{1.000000,1.000000,1.000000}%
\pgfsetstrokecolor{currentstroke}%
\pgfsetdash{}{0pt}%
\pgfpathmoveto{\pgfqpoint{1.327841in}{0.171373in}}%
\pgfpathlineto{\pgfqpoint{1.327841in}{0.175779in}}%
\pgfpathlineto{\pgfqpoint{1.327841in}{1.294809in}}%
\pgfpathlineto{\pgfqpoint{1.327841in}{1.299214in}}%
\pgfpathlineto{\pgfqpoint{1.440625in}{1.299214in}}%
\pgfpathlineto{\pgfqpoint{1.440625in}{1.294809in}}%
\pgfpathlineto{\pgfqpoint{1.440625in}{0.175779in}}%
\pgfpathlineto{\pgfqpoint{1.440625in}{0.171373in}}%
\pgfpathclose%
\pgfusepath{stroke,fill}%
\end{pgfscope}%
\begin{pgfscope}%
\definecolor{textcolor}{rgb}{0.150000,0.150000,0.150000}%
\pgfsetstrokecolor{textcolor}%
\pgfsetfillcolor{textcolor}%
\pgftext[x=1.537847in,y=0.123148in,left,base]{\color{textcolor}\sffamily\fontsize{10.000000}{12.000000}\selectfont 0}%
\end{pgfscope}%
\begin{pgfscope}%
\definecolor{textcolor}{rgb}{0.150000,0.150000,0.150000}%
\pgfsetstrokecolor{textcolor}%
\pgfsetfillcolor{textcolor}%
\pgftext[x=1.537847in,y=0.498970in,left,base]{\color{textcolor}\sffamily\fontsize{10.000000}{12.000000}\selectfont 100}%
\end{pgfscope}%
\begin{pgfscope}%
\definecolor{textcolor}{rgb}{0.150000,0.150000,0.150000}%
\pgfsetstrokecolor{textcolor}%
\pgfsetfillcolor{textcolor}%
\pgftext[x=1.537847in,y=0.874791in,left,base]{\color{textcolor}\sffamily\fontsize{10.000000}{12.000000}\selectfont 200}%
\end{pgfscope}%
\begin{pgfscope}%
\definecolor{textcolor}{rgb}{0.150000,0.150000,0.150000}%
\pgfsetstrokecolor{textcolor}%
\pgfsetfillcolor{textcolor}%
\pgftext[x=1.537847in,y=1.250613in,left,base]{\color{textcolor}\sffamily\fontsize{10.000000}{12.000000}\selectfont 300}%
\end{pgfscope}%
\begin{pgfscope}%
\pgfsys@transformshift{1.326667in}{0.170397in}%
\pgftext[left,bottom]{\pgfimage[interpolate=true,width=0.113333in,height=1.130000in]{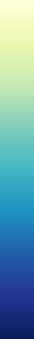}}%
\end{pgfscope}%
\begin{pgfscope}%
\pgfsetbuttcap%
\pgfsetmiterjoin%
\pgfsetlinewidth{1.003750pt}%
\definecolor{currentstroke}{rgb}{0.800000,0.800000,0.800000}%
\pgfsetstrokecolor{currentstroke}%
\pgfsetdash{}{0pt}%
\pgfpathmoveto{\pgfqpoint{1.327841in}{0.171373in}}%
\pgfpathlineto{\pgfqpoint{1.327841in}{0.175779in}}%
\pgfpathlineto{\pgfqpoint{1.327841in}{1.294809in}}%
\pgfpathlineto{\pgfqpoint{1.327841in}{1.299214in}}%
\pgfpathlineto{\pgfqpoint{1.440625in}{1.299214in}}%
\pgfpathlineto{\pgfqpoint{1.440625in}{1.294809in}}%
\pgfpathlineto{\pgfqpoint{1.440625in}{0.175779in}}%
\pgfpathlineto{\pgfqpoint{1.440625in}{0.171373in}}%
\pgfpathclose%
\pgfusepath{stroke}%
\end{pgfscope}%
\end{pgfpicture}%
\makeatother%
\endgroup%
		\subcaption{The countour of the cell body (orange) as segmented by HNCcorr.}
		\label{fig:algWindowSegm}
	\end{minipage}
%
	
	\begin{minipage}[t]{\textwidth}
		\vspace{0pt}
		\centering
		\input{figures/segmentSignal.pgf}
		\subcaption{The average intensity of the pixels in the segmented cell body over time. Each spike corresponds to an activation of the cell. }
		\label{fig:algSignal}
	\end{minipage}
	
	\caption{A visualization of the input to HNCcorr, the graph constructed by HNCcorr, and the resulting segmentation of the cell body for a window of $31 \times 31$ pixels from the Neurofinder 02.00 training dataset.}
	\label{fig:algorithm}
\end{figure}

\begin{figure}
	\centering
	\begin{minipage}[t]{0.48\textwidth}
		\vspace{0pt}
		\centering
		\includegraphics{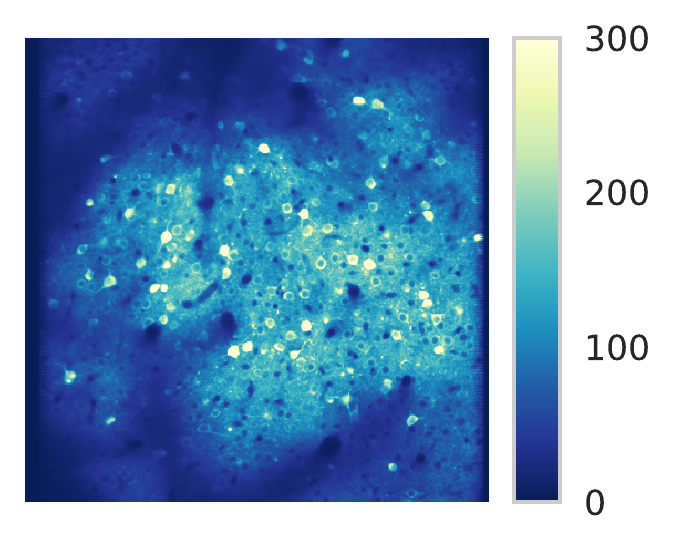}
		\subcaption{Overview image of the Neurofinder 02.00 training dataset. The image shows the average intensity of each pixel over time. }
		\label{fig:overviewWithout}
	\end{minipage}
	\hfill
	\begin{minipage}[t]{0.48\textwidth}
		\vspace{0pt}
		\centering
		\includegraphics{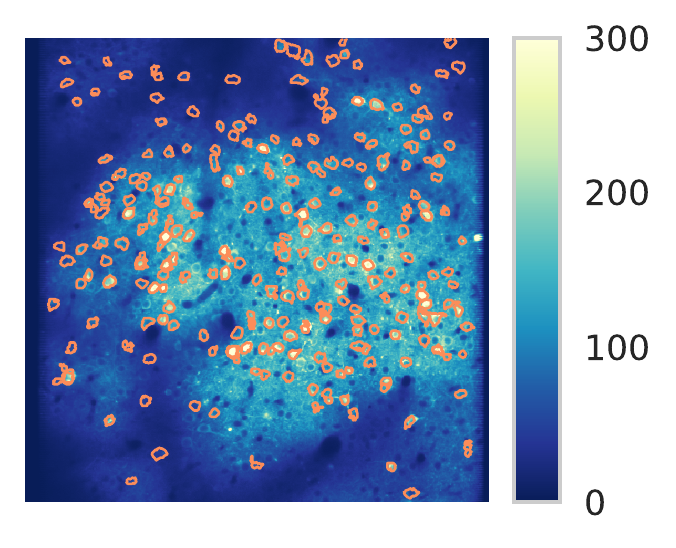}
		\subcaption{The orange shapes are the contours of the cell bodies as identified by HNCcorr for the Neurofinder 02.00 training dataset.}
		\label{fig:overviewWith}
	\end{minipage}
	\caption{Overview of the cell bodies identified by HNCcorr for the Neurofinder 02.00 training dataset.}
	\label{fig:overview}
\end{figure}

A visualization of a segmentation for a given seed is given in Figure \ref{fig:algorithm}. Figure \ref{fig:overview} shows an example of the output of the HNCcorr algorithm after processing all seeds in a calcium imaging movie.

	\section*{Experimental Results}
To evaluate the experimental performance of HNCcorr in detecting cells in calcium imaging movies, we compare against the best known algorithms to date.
Specifically, we test HNCcorr as well as the matrix factorization algorithms Suite2P~\cite{PacStrSch16} and CNMF~\cite{PneSouGao16} on the datasets of the Neurofinder public benchmark~\cite{NeuFinder16}. The Neurofinder benchmark is an initiative of the CodeNeuro collective of neuroscientists that encourages software tool development for neuroscience research.
Based on the Neurofinder benchmark, we show that HNCcorr outperforms the other algorithms with a relative improvement in average F1-score across datasets of at least 14 percent.

\begin{table}[h]
	\begin{center}
		\small
		\caption{Characteristics of the test datasets of the Neurofinder benchmark.}\label{tab:datasetsInfo}
		\begin{tabular}{llllll}
			\toprule
			Dataset name & Source of dataset & Length [s] & Frequency [hz] & Brain region &  Reference technique\\
			\cmidrule(r){1-1} \cmidrule(lr){2-2} \cmidrule(lr){3-3} \cmidrule(lr){4-4} \cmidrule(lr){5-5} \cmidrule(l){6-6}
			00.00.test & Svoboda Lab & 438 & 7.00 & vS1 & Anatomical markers \\
			00.01.test & Svoboda Lab & 458 & 7.00 & vS1 & Anatomical markers \\
			01.00.test & Hausser Lab & 300 & 7.50 & v1 & Human labeling \\
			01.01.test & Hausser Lab & 667 & 7.50 & v1 & Human labeling\\
			02.00.test & Svoboda Lab & 1000 & 8.00 & vS1 & Human labeling\\
			02.01.test & Svoboda Lab & 1000 & 8.00 & vS1 & Human labeling\\
			03.00.test & Losonczy Lab & 300 & 7.50 & dHPC CA1 & Human labeling\\
			04.00.test & Harvey Lab & 444 & 6.75 & PPC & Human labeling\\
			04.01.test & Harvey Lab & 1000 & 3.00 & PPC & Human labeling\\
			\bottomrule
		\end{tabular}
	\end{center}
\end{table}

The Neurofinder benchmark consists of nine testing datasets and nineteen training datasets.
Each dataset consists of a calcium imaging movie.
For the training datasets, a reference list of the spatial footprints of the cells in the movie, is provided with the dataset.
For the testing datasets, the reference list (i.e. the spatial footprints of the cells) is not disclosed.
The datasets have been contributed by various labs and have been recorded under different experimental conditions.
They differ in sample frequency, length of the movie, magnification, signal-to-noise ratio, region of the brain that was recorded, and how the footprints in the datasets were determined.
Table~\ref{tab:datasetsInfo} contains the characteristics of the testing datasets.

The datasets can be split into two groups based on the activity of the cells.
The reference list for the 00 and 03 dataset series contains many cells that have a weak or non-existent signal, whereas most cells have a detectable signal in the 01, 02, and 04 dataset series.
As a result, algorithms based on signal detection are expected to perform poorly on the 00 and 03 dataset series.

To evaluate an algorithm on the Neurofinder benchmark, the algorithm should identify the cells in each testing dataset.
The spatial footprints of the cells as detected are then submitted to Neurofinder.
The submission is then automatically evaluated by comparing it against the reference list of spatial footprints provided by the dataset contributors.
Since these lists are not publicly disclosed for the testing datasets, this guarantees an unbiased and fair evaluation.
Furthermore, this procedure removes the need for replicating the results of other algorithms.

A submission to Neurofinder is evaluated for each testing dataset based on metrics for detection quality and metrics for the segmentation quality\footnote{See the supplementary material for results on detection quality.}.  To measure the detection quality of the algorithm  we consider the following metrics,~\cite{NeuFinder16}:
\begin{description}[itemsep=0pt,partopsep=0pt]
  \item[Recall] The percentage of the cells in the undisclosed reference list of spatial footprints that are recovered by the algorithm.
  \item[Precision] The percentage of the cells identified by the algorithm that are also present in the undisclosed reference list.
  \item[F1-score] The harmonic mean of precision and recall. This is a standard metric in machine learning for evaluating the overall performance of an algorithm.
\end{description}
For the recall and precision metrics, a cell is considered identified if the center of mass of the spatial footprint as determined by the algorithm is within five pixels of the center of mass of the spatial footprint in the reference list.

\begin{figure}
		\centering
		\input{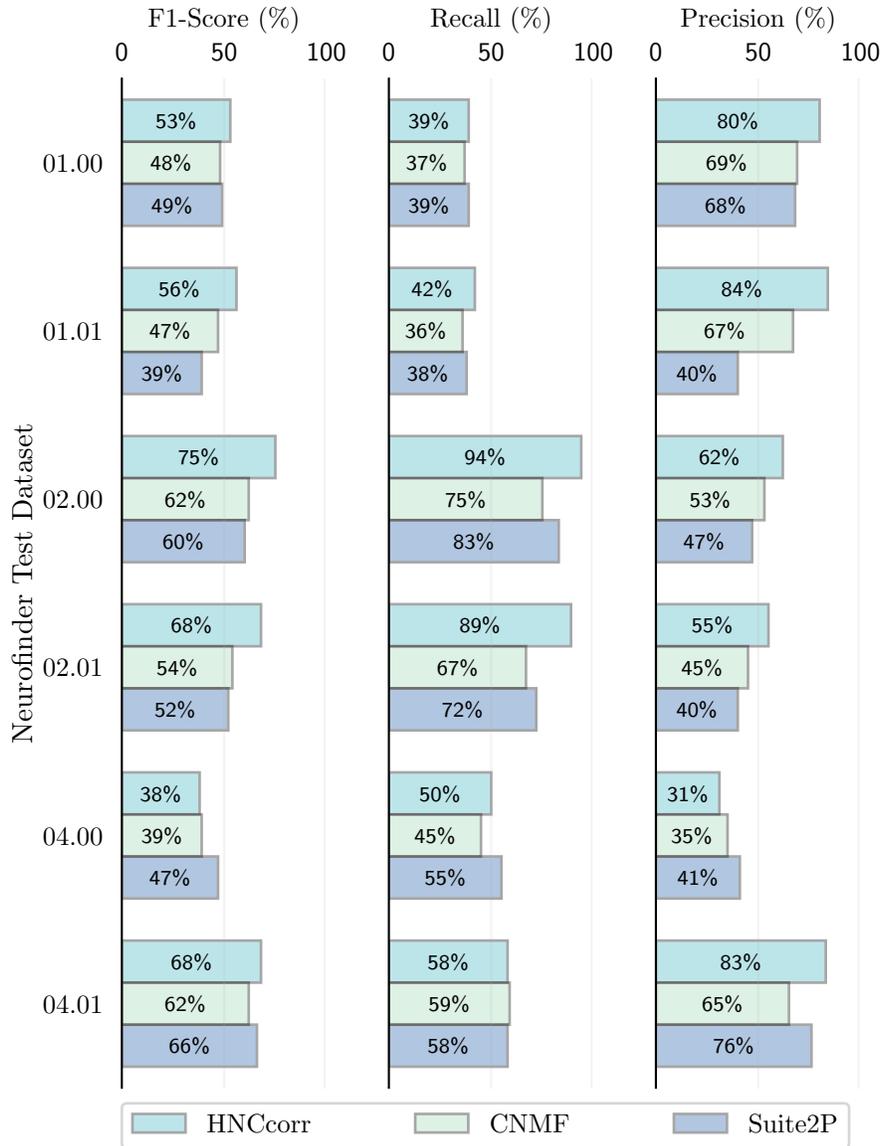}
		\caption{Experimental performance of the HNCcorr, CNMF, and Suite2P algorithms on the Neurofinder testing datasets of the 01, 02, 04 series.
		The 00 and 03 series have been excluded since these datasets are not appropriate for signal-based algorithms.
		For each of the listed metrics, higher scores are better.
	 	The data is taken from Neurofinder submissions \texttt{Suite2P} by \texttt{marius10p}, \texttt{CNMF\_PYTHON} by \texttt{CNMF}, and submission \texttt{HNCcorr} by \texttt{HNC}.}
	\label{fig:expResults}
\end{figure}

The experimental performance of the signal-based algorithms HNCcorr, CNMF, and Suite2P on the 01, 02, 04 dataset series is shown in Figure \ref{fig:expResults}. The implementation details for the HNCcorr submission are described in the supplementary material. Since the 00 and 03 dataset series are not appropriate for signal-based algorithms, these dataset series have been excluded. Results on all testing datasets are provided in the supplementary material.

Overall, the HNCcorr algorithm has superior performance across the datasets.
The HNCcorr achieves a 14 percent relative improvement in average F1-score compared to the Suite2P and CNMF algorithm.
The HNCcorr algorithm also achieves the highest F1-score for each of the datasets with the exception of the dataset 04.00.
This improvement is mainly due to an increase in precision across all datasets except 04.00.
All of the algorithms have a significantly lower precision for the 04.00 dataset as compared to the other datasets.
Based on analysis of the 04.00 training dataset, a possible explanation is that some cells with signal are identified by the algorithms but these cells are not in the reference listing.
Furthermore, the HNCcorr algorithm achieves a significant improvement in recall for the 02.00 and 02.01 datasets, resulting in near-perfect recall.

\begin{figure}
	\centering
	\input{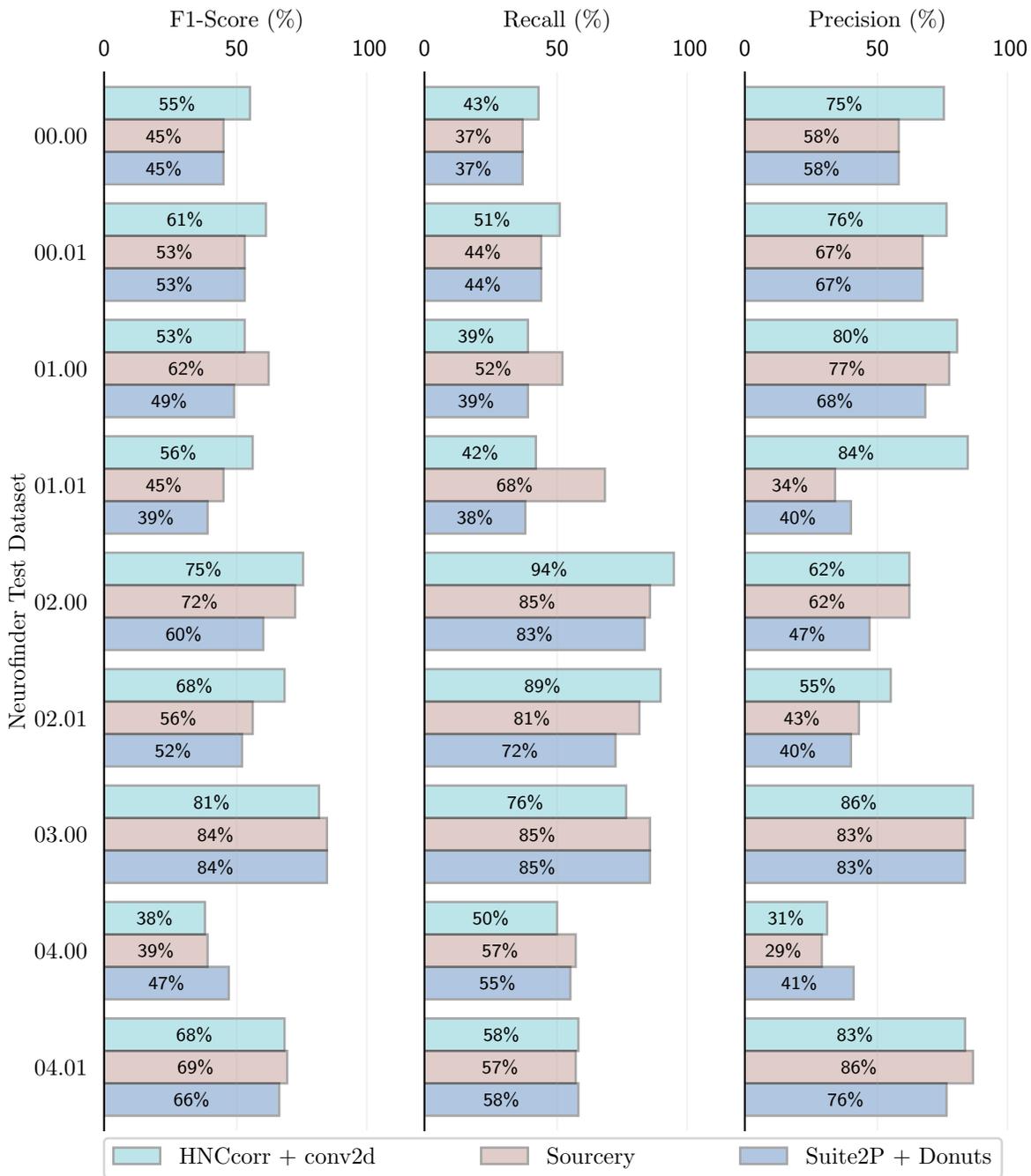}
	\vspace{-0.25cm}
	\caption{Experimental performance on all testing datasets of the three leading submissions of the Neurofinder benchmark as of January 2017.
		For each of the listed metrics, higher scores are better.
		The HNCcorr algorithm combined with the conv2d shape-based detection achieves the highest average F1-score.
		Both the \texttt{Suite2P} and the \texttt{Sourcery} submission use the Donuts algorithm~\cite{PacPacPet13} for the 00 and 03 dataset series and a version of the \texttt{Suite2P} algorithm~\cite{PacStrSch16}.
		The details of the \texttt{Suite2p} variant used for the \texttt{Sourcery} submission have not been released by the time of writing.
		For the HNCcorr + conv2d submission, the results on the 00 and 03 series differ from the \texttt{Conv2d} submission since the model was retrained by the authors of this paper.}
	\label{fig:expResultsCombined}
\end{figure}

A common strategy for the Neurofinder benchmark is to submit an ensemble of algorithms to mitigate the poor performance of the signal-based algorithms on the 00 and 03 dataset series.
Typically, the ensemble consists of a shape-based detection for the 00 and 03 series and a signal-based algorithm for the 01, 02, and 04 dataset series.
In Figure~\ref{fig:expResultsCombined}, we compare the results of HNCcorr combined with conv2d with two other ensembles of algorithms based on Suite2P.
These three submissions are leading on the Neurofinder leaderboard as of January 2017, and the ensemble of the HNCcorr algorithm and the conv2d shape-based algorithm achieves the highest score for the Neurofinder benchmark.

	\section*{Discussion}
The algorithm HNCcorr is a new algorithm for cell identification in calcium imaging movies. In contrast to previous methods, HNCcorr is not based on matrix factorization. Instead, the HNCcorr algorithm combines graph clustering based on combinatorial optimization with the use of correlations.  Here we discuss several characteristics of HNCcorr.

First, the distinguishing feature of HNCcorr is that its underlying optimization model, the HNC model, can be solved optimally with a combinatorial optimization algorithm.
To the best of our knowledge, all other techniques rely on heuristics for finding a solution to their model.
Solving a model optimally is important because the solution is fully determined by the model and does not depend on the solution technique.
This provides a transparent mapping from the model and the input data to the outcome.
As a result, unsatisfactory results can therefore be analyzed solely based on the model, without having to consider any uncertainty introduced by the solution technique.

Second, the HNCcorr algorithm has a different computational structure.
The key difference compared to other algorithms is that the algorithm detects the cell bodies of the neurons one at a time.
An immediate advantage is that multiple seeds can be evaluated simultaneously.
This allows for efficient parallelization of the algorithm.
Another characteristic of HNCcorr is that it has an oracle that filters segmentations that are not cells.
This is used since HNCcorr attempts to identify a cell based on any seed.
As demonstrated, a simple oracle based on the size of the segmentation is sufficient to attain state-of-the-art precision.
We believe that a more advanced oracle could significantly improve the precision of the algorithm, and we plan to investigate this further.
Another observation is that the HNCcorr algorithm scales linearly in the length of the movie whereas matrix factorization algorithms do not.

A critical parameter for the HNCcorr algorithm is the size of the window.
The window size sets the dimension of the correlation space and determines the computational cost of mapping the pixels to correlation space.
If this becomes a significant computational issue, then we plan to explore the following ideas:
One is to limit the computation of correlation only to the relevant similarities and set the remaining correlations to zero.
For example, one could use an unbiased estimator for the correlation based on a random subset of the frames and compute only those correlation that are significantly different from zero.
Another option is to limit the dimension of the correlation space by computing only the correlations with respect to a (possibly random) subset of the pixels.

\begin{figure}
	\centering
	\input{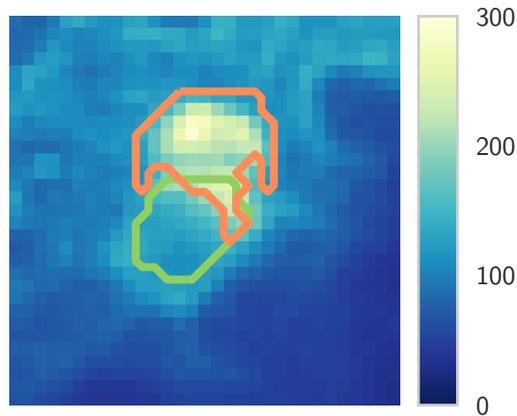}
	\caption{An example of two overlapping cells from the Neurofinder.02.00 training dataset. The red and green lines each mark the contour of a cell. The background image shows the average intensity of each pixel over time. }
	\label{fig:overlappingNeuron}
\end{figure}

An often stated advantage of matrix factorization based algorithms is that these algorithms can identify overlapping cells based on their signal.
Although the percentage of cells that overlap is typically small, the HNCcorr algorithm is also able to identify overlapping cells, see Figure \ref{fig:overlappingNeuron} for an example.
The key observation here is that pixels can be segmented more than once and, thus, belong to multiple cells.

Finally, a novel feature of HNCcorr is the use of correlation space.
Through the mapping to correlation space, HNCcorr is the first algorithm that employs correlation to attain state-of-the-art results.
As previously discussed in Figures \ref{fig:corrspace} and \ref{fig:corrspaceMat}, the success of employing correlation space is that closeness in this space is a strong indicator of whether pixels correspond to the same cells or background.

	\section*{Conclusion}
In this paper we present HNCcorr for identifying cells in calcium imaging movies.
HNCcorr segments cells one at a time by finding distinct sets of pixels that are nearly identical in correlation space.
The algorithm is based on a combinatorial optimization problem known as HNC that can be solved efficiently.
As a result, the algorithm guarantees to find an optimal solution to the underlying optimization problem.
Experimentally, HNCcorr has superior performance compared to matrix factorization based algorithms that have been considered state-of-the-art.
For cells segmented based on activity, HNCcorr achieves a relative improvement of at least 14 percent in average F1-score compared to CNMF and Suite2P.
Combined with the Conv2d shape-based detection algorithm, HNCcorr achieves the best known results to date in the Neurofinder benchmark.

For further research, we plan to adapt HNCcorr to calcium imaging movies collected with light-field microscopy~\cite{PrevYooHof14} as well as for the detection of subcellular components (e.g. dendrites).
We also plan to explore a more powerful oracle that decides whether a segmentation corresponds to a cell which could potentially lead to a significant improvement in the performance of the algorithm.
Finally, we plan to investigate other domains where HNCcorr could be of value.

	\section*{Acknowledgments}
The authors would like to thank Christopher Harvey, Selmaan Chettih, and Matthias Minderer from Harvard Medical School for their guidance, the data they provided, and for the valuable discussions we have had.

	\bibliography{references/references}

  \clearpage
\section*{Supplementary Material}

\subsection*{Implementation details}
HNCcorr was implemented in Matlab R2015a. Here we describe the relevant implementation details.

A new Python version of the code has been made available on GitHub: \url{https://github.com/quic0/HNCcorr}
\begin{description}
	\item[Preprocessing:] The datasets were preprocessed by averaging every ten frames into a single frame to reduce the noise. Each preprocessed dataset is stored in HDF5 format for efficient data access.
	\item[Seed selection:] 
	To identify possibly effective seeds, we use the procedure described in Algorithm \ref{alg:HNCcorr}.
	\begin{algorithm}[H]
		\caption{HNCcorr seed selection and outer loop.}
		\label{alg:HNCcorr} 
		\begin{algorithmic}
			\State \emph{seeds} $\gets \emptyset$
			\State \emph{segmentations} $\gets \emptyset$
			\State 
			\State \texttt{\footnotesize// segmentedPixels contains the pixels that have been segmented at least once}
			\State \emph{segmentedPixels} $\gets \emptyset$
			\State
			\State \texttt{\footnotesize// Select initial set of seeds}
			\State Split the pixels into a grid of $5\times5$ pixels starting at north-west pixel.
			\For{ each grid block $b$}
				\State Select pixel $i^*$ from grid block $b$ with highest mean correlation between pixel $i \in b$ and 
				\State all pixels in the $5 \times 5$ neighborhood centered at pixel $i$.
				\State \emph{seeds} $\gets$ \emph{seeds} $\cup$ $ \{ i^* \}$
			\EndFor
			\State
			\State Keep 40\% of \emph{seeds} with highest mean correlation to the neighborhood and sort in descending order 
			\State according to their mean correlation. Discard the remaining seeds.
			\State
			\State \texttt{\footnotesize// Attempt segmentation seed}
			\For{ each seed $s$ in seeds}
				\If{ $s \notin$ \emph{segmentedPixels}}
					\State \texttt{\footnotesize/* \textsc{HNCcorrSingleSegmentation} is the main subroutine of HNCcorr that takes in a seed and}
					\State \texttt{\footnotesize returns a segmentation (possibly empty). */}
					\State \emph{segmentation} $\gets$  \Call{HNCcorrSingleSegmentation}{\emph{seed}}
					\If{ \emph{segmentation} $\neq \emptyset$ }
						\State \emph{segmentedPixels} $\gets$ \emph{segmentedPixels} $\cup$ \emph{segmentation}
						\State \emph{segmentations} $\gets$ \emph{segmentations} $\cup$ $\{ \text{\emph{segmentation}}\}$
					\EndIf
				\EndIf
			\EndFor
		\end{algorithmic}
	\end{algorithm}
\clearpage
	\item[Parameters settings:]
	For sparse computation we use three dimensions ($p=3$) for the low-dimensional space and a grid resolution of $\kappa = 25$.
	For the negative seeds, we select a set of nine pixels uniformly distributed from a large circle.
	~\\
	The remaining parameters depend on the dataset and are defined as follows:
	\begin{center}
		\captionof{table}{Parameters for the HNCcorr implementation as used in the experimental results.}
		\begin{tabular}{ccccccc}
			\toprule
			&&Radius circle & Seed size &  Oracle & Oracle & Oracle \\
			Dataset & Window size & negative seeds & superpixel & lower bound & upper bound & expected size \\
			\cmidrule(r){1-1} \cmidrule(lr){2-2} \cmidrule(lr){3-3} \cmidrule(lr){4-4} \cmidrule(lr){5-5} \cmidrule(lr){6-6} \cmidrule(l){7-7}
			00.00.test & $31 \times 31$ & $10$  pixels& $ 5 \times 5$ &  $40$ pixels & $150$ pixels & $60$ pixels \\ 
			00.01.test & $31 \times 31$ & $10$ pixels& $ 5 \times 5$ &  $40$ pixels & $150$ pixels & $65$ pixels \\ 
			01.00.test & $41 \times 41$ & $14$ pixels&$ 5 \times 5$ &  $40$ pixels & $380$ pixels & $170$ pixels \\ 
			01.01.test & $41 \times 41$ & $14$ pixels&$ 5 \times 5$ &  $40$ pixels & $380$ pixels & $170$ pixels \\ 
			02.00.test & $31 \times 31$ & $10$ pixels&$ 1 \times 1$ &  $40$ pixels & $200$ pixels & $80$ pixels \\ 
			02.01.test & $31 \times 31$ & $10$ pixels&$ 1 \times 1$ &  $40$ pixels & $200$ pixels & $80$ pixels \\ 
			03.00.test & $41 \times 41$ & $14$ pixels &$ 5 \times 5$ &  $40$ pixels & $300$ pixels & $120$ pixels \\ 
			04.00.test & $31 \times 31$ & $10$ pixels&$ 3 \times 3$ &  $50$ pixels & $190$ pixels & $90$ pixels \\ 
			04.01.test & $41 \times 41$ & $14$ pixels&$ 3 \times 3$ &  $50$ pixels & $370$ pixels & $140$ pixels \\ 
			\bottomrule
		\end{tabular}
	\end{center}
\end{description}

\clearpage
\subsection*{Neurofinder Benchmark results}
The tables below show the performance of the algorithms on all datasets for both the cell detection and segmentation quality metrics. The metrics for cell detection have been defined in the experimental results section. To measure the segmentation quality, the Neurofinder benchmark uses the metrics inclusion and exclusion~\cite{NeuFinder16}.

\subsubsection*{Metrics segmentation quality}
Let $A$ be the set of pixels in the spatial footprint of a cell for the reference labeling and $B$ be the set of pixels of the spatial footprint of a cell for the algorithm.
Then, the metrics inclusion and exclusion are defined as follows:
\begin{description}[itemsep=0pt,partopsep=0pt]
	\item[Inclusion] Average of $\frac{\left|A \cap B\right|}{\left| A \right|}$ across all cells correctly identified by the algorithm.
	\item[Exclusion] Average of $\frac{\left|A \cap B\right|}{\left| B \right|}$ across all cells correctly identified by the algorithm.
\end{description}
These metrics are of lower importance since the critical task is finding the location of a cell. 
It may still be possible to extract the signal of the cell even if the shape is not detected perfectly.
Also, the spatial footprint of the cell in the reference labeling is not always fine-tuned.
 
\subsubsection*{Results}
For each of the listed metrics, higher scores are better. 
\begin{table}[h]
	\centering
   	\caption{Experimental performance of the HNCcorr, CNMF, and Suite2P algorithms on all Neurofinder testing datasets.
   	The data is taken from Neurofinder submissions \texttt{Suite2P} by \texttt{marius10p}, \texttt{CNMF\_PYTHON} by \texttt{CNMF}, and submission \texttt{HNCcorr} by \texttt{HNC}.}
   \label{tab:resultsNeurofinder}
   	\begin{tabular}{ccrrrrrrrrr}
   		\toprule
  		\multirow{2}{*}{\bf Metric} & \multirow{2}{*}{\bf Algorithm} & \multicolumn{9}{c}{ \bf Datasets} \\
	   \cmidrule(lr){3-11}
       & & 00.00 & 00.01 & 01.00 & 01.01 & 02.00 & 02.01 & 03.00 & 04.00 & 04.01\\
   		\cmidrule(r){1-1} \cmidrule(lr){2-2} \cmidrule(r){1-1} \cmidrule(lr){2-2} \cmidrule(l){3-11}
       \multirow{3}{*}{F1} & HNCcorr & 29 & 33 & 53 & 56 & 75 & 68 & 23 & 38 & 68\\
                           & CNMF    & 28 & 35 & 48 & 47 & 62 & 54 & 28 & 39 & 62\\
                           & Suite2P & 32 & 39 & 49 & 39 & 60 & 52 & 23 & 47 & 66\\
       \cmidrule(r){1-1} \cmidrule(lr){2-2} \cmidrule(r){1-1} \cmidrule(lr){2-2} \cmidrule(l){3-11}
       \multirow{3}{*}{Recall} & HNCcorr     & 19 & 25 & 39 & 42 & 94 & 89 & 15 & 50 & 58\\
                               & CNMF    & 18 & 24 & 37 & 36 & 75  & 67 & 20 & 45 & 59\\
                               & Suite2P & 20 & 27 & 39 & 38 & 83  & 72 & 15 & 55 & 58\\
       \cmidrule(r){1-1} \cmidrule(lr){2-2} \cmidrule(r){1-1} \cmidrule(lr){2-2} \cmidrule(l){3-11}
       \multirow{3}{*}{Precision} & HNCcorr     & 62 & 52 & 80 & 84 & 62 & 55 & 47 & 31 & 83\\
                               & CNMF   & 69 & 65 & 69 & 67 & 53 & 45 & 49 & 35 & 65\\
                               & Suite2P & 80 & 72 & 68 & 40 & 47 & 40 & 57 & 41 & 76\\
      \cmidrule(r){1-1} \cmidrule(lr){2-2} \cmidrule(r){1-1} \cmidrule(lr){2-2} \cmidrule(l){3-11}
       \multirow{3}{*}{Inclusion} & HNCcorr     & 63 & 65 & 58 & 56 & 78 & 76 & 55 & 68 & 74\\
                               & CNMF    & 73 & 79 & 56 & 51 & 79 & 80 & 64 & 70 & 77\\
                               & Suite2P & 72 & 78 & 56 & 45 & 78 & 77 & 78 & 68 & 86\\
       \cmidrule(r){1-1} \cmidrule(lr){2-2} \cmidrule(r){1-1} \cmidrule(lr){2-2} \cmidrule(l){3-11}
       \multirow{3}{*}{Exclusion} & HNCcorr     & 69 & 65 & 93 & 94 & 84 & 85 & 35 & 90 & 87\\
                               & CNMF    & 66 & 68 & 96 & 97 & 87 & 87 & 43 & 90 & 88\\
                               & Suite2P & 65 & 64 & 92 & 95 & 81 & 82 & 38 & 92 & 83\\
       \bottomrule
   	\end{tabular}
   
\end{table}
\begin{table}
	\centering
    \caption{Experimental performance on all testing datasets of the three leading submissions of the Neurofinder benchmark as of January 2017. 
   	Both the \texttt{Suite2P} and the \texttt{Sourcery} submission use the Donuts algorithm~\cite{PacPacPet13} for the 00 and 03 dataset series and a version of the \texttt{Suite2P} algorithm~\cite{PacStrSch16}. }
   \label{tab:resultsNeurofinderCombined}
   \begin{tabular}{ccrrrrrrrrr}
   	\toprule
   	\multirow{2}{*}{\bf Metric} & \multirow{2}{*}{\bf Algorithm} & \multicolumn{9}{c}{ \bf Datasets} \\
   	\cmidrule(l){3-11}
   	& & 00.00 & 00.01 & 01.00 & 01.01 & 02.00 & 02.01 & 03.00 & 04.00 & 04.01\\
   	\cmidrule(r){1-1} \cmidrule(lr){2-2} \cmidrule(l){3-11}
   	\multirow{3}{*}{F1} & HNCcorr + conv2d        & 55 & 61 & 53 & 56 & 75 & 68 & 82 & 38 & 68\\
   	& Sourcery                & 45 & 53 & 62 & 45 & 72 & 56 & 84 & 39 & 69\\
   	& Suite2P + Donuts        & 45 & 53 & 49 & 39 & 60 & 52 & 84 & 47 & 66\\
   	\cmidrule(r){1-1} \cmidrule(lr){2-2} \cmidrule(r){1-1} \cmidrule(lr){2-2} \cmidrule(l){3-11}
   	\multirow{3}{*}{Recall} & HNCcorr + conv2d    & 43 & 51 & 39 & 42 & 94 & 89 & 76 & 50 & 58\\
   	& Sourcery            & 37 & 44 & 52 & 68 & 85 & 81 & 85 & 57 & 57\\
   	& Suite2P + Donuts    & 37 & 44 & 39 & 38 & 83 & 72 & 85 & 55 & 58\\
   	\cmidrule(r){1-1} \cmidrule(lr){2-2} \cmidrule(r){1-1} \cmidrule(lr){2-2} \cmidrule(l){3-11}
   	\multirow{3}{*}{Precision} & HNCcorr + conv2d    & 75 & 76 & 80 & 84 & 62 & 55 & 86 & 31 & 83\\
   	& Sourcery            & 58 & 67 & 77 & 34 & 62 & 43 & 83 & 29 & 86\\
   	& Suite2P + Donuts    & 58 & 67 & 68 & 40 & 47 & 83 & 57 & 41 & 76\\
   	\cmidrule(r){1-1} \cmidrule(lr){2-2} \cmidrule(r){1-1} \cmidrule(lr){2-2} \cmidrule(l){3-11}
   	\multirow{3}{*}{Inclusion} & HNCcorr  + conv2d   & 55 & 56 & 58 & 56 & 78 & 76 & 75 & 68 & 74\\
   	& Sourcery            & 98 & 98 & 81 & 72 & 96 & 95 & 100 & 92 & 98\\
   	& Suite2P + Donuts    & 98 & 98 & 56 & 45 & 78 & 77 & 100 & 68 & 86\\
   	\cmidrule(r){1-1} \cmidrule(lr){2-2} \cmidrule(r){1-1} \cmidrule(lr){2-2} \cmidrule(l){3-11}
   	\multirow{3}{*}{Exclusion} & HNCcorr + conv2d    & 78 & 77 & 93 & 94 & 84 & 79 & 35 & 90 & 87\\
   	& Sourcery            & 36 & 35 & 81 & 84 & 56 & 58 & 31 & 73 & 63\\
   	& Suite2P + Donuts    & 36 & 35 & 92 & 95 & 81 & 82 & 31 & 92 & 83\\
   	\bottomrule
   \end{tabular}
\end{table}

\end{document}